\documentclass{article}
\usepackage{amsfonts}
\usepackage{mathrsfs}
\usepackage{amsmath}
\usepackage{verbatim}
\usepackage{amssymb}
\usepackage{amsfonts,amsmath,latexsym,amssymb,txfonts,bm}
\usepackage[american]{babel}
\usepackage[american]{babel}

\makeatletter
\renewcommand{\@evenhead}{\raisebox{0pt}[\headheight][0pt]{\vbox{\hbox
to \textwidth{\thepage\hfil\strut\textit{\leftmark}}\hrule}}}
\renewcommand{\@oddhead}{\raisebox{0pt}[\headheight][0pt]{\vbox{\hbox
to \textwidth{\textit{\rightmark}\hfil\strut\thepage}\hrule}}}
\makeatother

\renewcommand\theequation{\thesection.\arabic{equation}}

\newcommand{\bra}[1]{\langle{#1}}
\newcommand{\ket}[1]{{#1}\rangle}

\def\II{{\mathbb I}}
\def\RR{{\mathbb R}}

\def\tr{\mathrm{ tr\,}}

\def\End{\mathrm{End}}

\def\be{\begin{equation}}
\def\ee{\end{equation}}
\def\bea{\begin{eqnarray}}
\def\eea{\end{eqnarray}}
\def\bed{\begin{definition}{\ }}
\def\eed{\end{definition}}
\def\bd{\begin{description}}
\def\ed{\end{description}}
\def\bc{\begin{center}}
\def\ec{\end{center}}

\newtheorem{definition}{Definition}

\def\sideremark#1{\ifvmode\leavevmode\fi\vadjust{\vbox to0pt{\vss
\hbox to 0pt{\hskip\hsize\hskip1em
\vbox{\hsize2cm\tiny\raggedright\pretolerance10000
\noindent #1\hfill}\hss}\vbox to8pt{\vfil}\vss}}}

\begin{document}

\begin{titlepage}
\thispagestyle{empty} \null

\hspace*{50truemm}{\hrulefill}\par\vskip-4truemm\par
\hspace*{50truemm}{\hrulefill}\par\vskip5mm\par
\hspace*{50truemm}{{\large\sc New Mexico Tech {\rm
(\today)}}}\vskip4mm\par
\hspace*{50truemm}{\hrulefill}\par\vskip-4truemm\par
\hspace*{50truemm}{\hrulefill}
\par
\bigskip
\bigskip
%\par
%\hspace*{50truemm}{\LARGE\textbf{\textsf{DRAFT}}}
%\par
\vspace{3cm}
\centerline{\huge\bf Non-commutative Corrections}
\bigskip
\centerline{\huge\bf in Spectral Matrix Gravity}
\bigskip
%\bigskip
%\bigskip
\bigskip
\centerline{\Large\bf Guglielmo Fucci and Ivan G. Avramidi}
\bigskip
\centerline{\it New Mexico Institute of Mining and Technology}
\centerline{\it Socorro, NM 87801, USA}
\centerline{\it E-mail: iavramid@nmt.edu, gfucci@nmt.edu}
\bigskip
\medskip
\vfill

{\narrower
\par
% ABSTRACT
We study a non-commutative deformation of general relativity based
on spectral invariants of a partial differential operator acting
on sections of a vector bundle over a smooth manifold. We compute
the first non-commutative corrections to Einstein equations in the
weak deformation limit and analyze the spectrum of the theory.
Related topics are discussed as well.

\par}
\vfill

%{\vbox{ \hrule \vspace{3pt} \hfil {\scriptsize\textit{
%\hfill\hfill\jobname.tex; \today; \timenow; p. \theoutputpage}}
%\hfil }}

\end{titlepage}

%=================================================================
\section{Introduction}
\setcounter{equation}0

The basic concept of general relativity is the space-time, which is
a smooth manifold with a globally hyperbolic metric $g_{\mu\nu}$.
As it is well known that the dynamics of the metric is described by
the Einstein-Hilbert action functional (with cosmological constant)
\begin{equation}
\label{2d}
S=\frac{1}{16\pi G}\int\limits_{M}dx\;g^{1/2}(R-2\Lambda)\;,
\end{equation}
where $dx$ denotes the standard Lebesgue measure on $M$,
$g=\det g_{\mu\nu}$, $R$ is the scalar curvature of the metric,
$G$ is the Newtonian gravitational constant and
$\Lambda$ is the cosmological constant.

It is a rather old idea that the metric can be described by the
properties of light propagating in the spacetime (that is, by
studying the properties of the wave equation) and the action of
general relativity can be `induced' by spectral invariants of a
second order partial differential operator of Laplace type (for
more details, see \cite{avramidi04}). In a series of recent papers
\cite{avramidi04,avramidi03,avramidi04b} a new theory of gravity
has been developed, called Matrix Gravity, in which the main idea
is to describe the gravitational field by studying the propagation
of fields with some internal structure rather than light. This
idea follows from the fact that at a more fundamental level (short
distances or high energies) the role of the photon could be played
by a multiplet of gauge fields. In the papers
\cite{avramidi04,avramidi04b} the action was constructed by
directly generalizing the language of Riemannian geometry to the
non-commutative case. In \cite{fucci08} we studied the
non-commutative limit of the theory proposed in
\cite{avramidi04,avramidi04b}. It turned out that this deformation
was non-unique. That is why, in \cite{avramidi03} a new model,
called Spectral Matrix Gravity, was developed, which is based on
spectral invariants of non-Laplace type partial differential
operators. In the present paper we study the weak deformation
limit in Spectral Matrix Gravity.

We would like to stress, at this point, that our approach to
deform General Relativity is \emph{different} from the more standard approach
of noncommutative extension of gravity on noncommutative spaces.
In flat space one usually introduces
noncommutative coordinates satisfying the commutation relations
\begin{equation}
[x^{\mu},x^{\nu}]=\theta^{\mu\nu}\;,
\end{equation}
where $\theta^{\mu\nu}$ is a real constant anti-symmetric matrix,
and one replaces the standard algebra of functions with the non-commutative
algebra with the Moyal star product
\begin{equation}
f(x)\;\star\;g(x)=\exp\left(\frac{i}{2}\theta^{\mu\nu}\frac{\partial}{\partial y^{\mu}}\frac{\partial}{\partial z^{\nu}}\right)f(x+y)g(x+z)\Big|_{y=z=0}\;.
\end{equation}
An extensive review of different realizations of gravity in the framework
of noncommutative geometry, especially in connection with string theory, can be found in \cite{szabo06}.
An extension of the star product and noncommutativity from flat to curved
space-time can be found in \cite{aschieri05,harikumar06}.

We list below the most relevant differences
between the two approaches,
a more detailed and extensive discussion can be found in
\cite{avramidi04,avramidi04b,avramidi08,fucci08}.

The biggest problem with the curved manifolds is the nature of the object
$\theta^{\mu\nu}$. All these models are defined, strictly speaking, only in
perturbation theory in the deformation parameter, that is, one takes
$\theta^{\mu\nu}$ as a {\it formal} parameter and considers {\it formal power series} in $\theta^{\mu\nu}$. 
In our approach to Matrix Gravity the deformation parameters are 
not formal and the theory is defined for all finite
values of the deformation
parameter.

In the standard non-commutative approach the coordinates themselves are non-commutative. This condition 
raises the questions of whether the space-time has the structure of a manifold, and how one can 
define analysis on such spaces. Moreover, one needs a way to relate the non-commutative coordinates 
with the usual (commutative) coordinates. 
In Matrix Gravity we do not have non-com\-mu\-ta\-tive coordinates.
Our space-time is a proper {\it smooth manifold} with the standard analysis defined on it.

In non-commutative geometry approach the deformation parameter $\theta^{\mu\nu}$ is {\it non-dynamical}, 
therefore there are no dynamical equations for it. This poses the question of what kind of physical, 
or mathematical, conditions can be used in order to determine it. In addition, in many models of non-commutative gravity
(as in \cite{aschieri05}) $\theta^{\mu\nu}$ is a 
\emph{non-tensorial} object 
which makes it dependent on the choice of the system of coordinates.
This feature leads the theory to be not invariant under the usual group of diffeomorphisms. 
In \cite{aschieri05} the authors construct all the relevant geometric
quantities like connection, curvature, etc. in terms of the non-commutative
deformation parameter. In this framework they obtain an
expansion of these quantities and of
the action up to second order. In their approach the object
$\theta^{\mu\nu}$ is a constant anti-symmetric matrix ({\it not a tensor}).
This violates 
the usual diffeomorphism invariance, Lorentz invariance, etc. 
for which there exist very strict experimental bounds.
The main result in \cite{aschieri05} is the derivation of the deformed Einstein equations.
The zero-order part (Einstein) is diffeomorphism-invariant, and the
corrections (quadratic in theta) are not. Therefore the theory contains some preferred 
system of coordinates and its whole content depends on it. 
Of course, the theory needs 
to justify the choice of such system of coordinates.

In our approach, we do not introduce any non-tensorial objects. As a result our theory is
{\it diffeomorphism-invariant}. So, there are no problems related to the violation
of Lorenz invariance, etc. and there are no preferred systems of coordinates. Moreover,
the non-commutative part of the metric in our approach is {\it dynamical}.
We have non-commutative Einstein equations for it. The goal of this paper is,
in particular, to derive these dynamical equations in the perturbation
theory.

In \cite{harikumar06} the authors
assume that $\theta^{\mu\nu}$
is a covariantly constant tensor.
But then, there are strict
algebraic constraints on the Riemann curvature tensor of the commutative
metric (obtained by a commutator of second covariant derivatives). 
In Matrix Gravity such algebraic constraints are absent---
the commutative metric is arbitrary.

In the usual approach of non-commutative geometry
(as in, for example, \cite{aschieri05}), 
when one defines the affine connection,
the covariant derivative, the curvature and the torsion
the {\it ordering of factors is not unique}. 
There is no natural reason why
one should prefer one ordering over the other. That is, the connection
coefficients can be placed on the left, or on the right, (or one
could symmetrize over
these two possibilities) from the object of
differentiation.
Another aspect of the ordering problem is the fact that
there is no unique way to raise and lower indices. One
can act with the metric from the left or from the right. 
Spectral Matrix Gravity, instead,
is pretty much {\it unique}. There is no need to define
the affine connection, the covariant derivative, the curvature and the
torsion. There is no ordering problem.

The definition of a ``measure", in standard non-commutative geometry, as a star determinant (as in \cite{aschieri05})
does not
guarantee its positivity. It only guarantees the
positivity in the zero order of the perturbation theory.
With our definition, the measure is positive even in strongly
non-com\-mu\-ta\-tive regime.

Moreover, the Moyal star product is non-local which makes the whole theory
{\it non-local} with possible unitarity problems. 
In Spectral Matrix Gravity approach the action functional is 
a usual local functional of sigma-model
type (like General Relativity, but with additional 
non-commutative degrees of freedom).
There may be problems with the renormalizability 
(which requires further study)
but not with unitarity.

We consider the space-time as a smooth manifold in which the gravitational
field is described by propagation of gauge fields rather than light. Therefore
one is naturally led to consider non-Laplace type partial differential operator
for which the leading symbol is now a matrix. One can evaluate the heat kernel
asymptotic coefficients for such operator and construct, from the first two,
the action for Matrix Gravity as in (\ref{2}). Similar calculations have been
performed in noncommutative geometry regarding heat kernel asymptotics
expansion. In \cite{sasakura04,sasakura05} the author evaluates the relevant
geometric quantities from an approximate power expansion of the trace of the
heat kernel for a Laplace operator on a compact fuzzy space. In
\cite{vassilevich} the author studies the quantization of noncommutative
gravity in two dimensions by considering a noncommutative deformation (using
the Moyal product) of the Jackiw-Teitelboim model for gravity. In this case the
path integral can be evaluated exactly and the operator for the quantum
fluctuations can be found. Once the operator is known one can study the first
two heat kernel asymptotic coefficients and obtain information about the
conformal anomaly and the Polyakov action. Our analysis is rather different
because we consider a standard smooth manifold manifold and a non-Laplace
operator with matrix valued leading symbol without considering Moyal product.

For the sake of completeness we describe, briefly, the framework
of Matrix Gravity (a more detailed discussion can be found in
\cite{avramidi04,avramidi04b,avramidi03}). Let us consider a
smooth compact orientable $n$-dimensional manifold $M$ without
boundary. Let $V[\omega]$ be a $N$-dimensional vector bundle over
$M$ of densities  of weight $\omega$ and $C^{\infty}(V[\omega])$
be the space of smooth sections of the vector bundle $V[\omega]$.
Locally, a section $\varphi$ of the vector bundle $V$ is described
by the $N$-vector $\varphi^A$. Let $\End(V)$ be the bundle of
endomorphisms of the vector bundle $V$; a section $Q$ of the
endomorphism bundle is represented locally by the matrix
$Q^A{}_B$. Next, we assume that the vector bundle $V[\omega]$ is
endowed with an Hermitian metric $G$ of weight $(1-2\omega)$
defining the fiber inner product $\bra{\phi}
,\ket{\psi}=\phi^{\dag}G\psi$. This enables one to define the
$L^2$-product of two sections of $V$  by
\begin{equation}\label{1}
(\phi,\psi)_{L^2}
=\int\limits_{M}dx\; \bra{\phi(x)} ,\ket{\psi(x)}\;,
\end{equation}
 The
completion of the space $C^{\infty}(V[\omega])$ in the
corresponding norm
defines the Hilbert space ${L}^{2}(V[\omega])$.

Now, let us consider a second order partial differential
operator
\begin{displaymath}
L:C^{\infty}(V[\omega])
%\longrightarrow
\to
C^{\infty}(V[\omega])\;,
\end{displaymath}
acting on smooth sections of the vector bundle $V[\omega]$,
with the endomorphism-valued coefficient functions
\begin{equation}
\label{1a}
L=-a^{\mu\nu}(x)\partial_{\mu}\partial_{\nu}
+b^{\mu}(x)\partial_{\mu}+c(x)\;.
\end{equation}
The leading symbol of the operator $L$ is defined as
\begin{equation}
\label{2a}
\sigma_L(L;x,\xi)=a^{\mu\nu}(x)\xi_{\mu}\xi_{\nu}\;,
\end{equation}
where $\xi_{\mu}$ is a cotangent vector at the point
$x$.

We consider two cases: the Euclidean case when the operator $L$ is
elliptic, and the pseudo-Euclidean case when the operator $L$ is
hyperbolic. These cases are related by an analytical continuation
and, for the sake of simplicity, we will restrict in our
calculation to the elliptic case. The formulas for the hyperbolic
case look exactly the same (for more details, see
\cite{avramidi04,avramidi04b,avramidi03}). The operator $L$ is
elliptic if the leading symbol $\sigma_L(L;x,\xi)$ is
non-degenerate for any $x$ and any $\xi\ne 0$. The operator $L$ is
of Laplace type if the leading symbol is essentially scalar, that
is,
\begin{equation}\label{2b}
\sigma(L;x,\xi)=\mathbb{I}g^{\mu\nu}(x)\xi_{\mu}\xi_{\nu}\;,
\end{equation}
where $\mathbb{I}$ is the identity matrix.

The geometric invariants we need can be derived
from the spectrum of the operator $L$.
The ${L}^{2}$-trace of the heat semigroup,
$\exp(-tL)$, usually called the heat trace, contains all the
information about the spectrum of the operator
$L$. It is well known that the
asymptotic expansion of the heat trace as $t\rightarrow 0^{+}$ has
the form
%============
\cite{gilkey95}
%============
\begin{equation}\label{vd}
\textrm{Tr}_{{L}^{2}}\;\exp(-tL)
\sim(4\pi)^{-\frac{n}{2}}\sum_{k=0}^{\infty}t^{k-\frac{n}{2}}A_{k}\;,
\end{equation}
where the coefficients $A_{k}$ are called the global heat kernel
coefficients, which are, of course, spectral invariants
of the operator $L$. They have the form
\begin{equation}\label{vd1}
A_{k}=\int\limits_{M}dx\;\;\textrm{tr}_{V} a_{k}\;,
\end{equation}
where $\textrm{tr}_{V}$ is the fiber trace on
the vector bundle $V$ and the coefficients $a_{k}$ are some
endomorphism-valued densities of weight one, called the local heat
kernel coefficients. It this the coefficients $A_0$ and $A_1$
that we study in this paper.

The action of Spectral Matrix Gravity proposed in
\cite{avramidi04b} is a linear combination of the heat kernel
coefficients $A_{0}$ and $A_{1}$, more precisely,
\begin{equation}\label{2}
S=\frac{1}{16\pi GN}\left[6A_{1}-2\Lambda
A_{0}\right]\;.
\end{equation}
We would like to point out here that the above action can be also thought of
as
a particular case of the Spectral Action Principle introduced in the framework
of noncommutative geometry in \cite{connes} and \cite{chamseddine93}.

For
the Laplace operator, $L=-\Delta$,
the heat kernel coefficients are well known
\begin{equation}\label{2c}
a_{0}=\II\;,\qquad a_{1}=\frac{1}{6}R\,\II\;.
\end{equation}
In this case, the action of Spectral Matrix Gravity reduces to the
standard Einstein-Hilbert action (\ref{2d}).

We would like to stress, here, that we are interested, in this
paper, in a much more complicated general case of an arbitrary
non-Laplace type operator (with a non-scalar leading symbol). In
this case there is no preferred Riemannian metric and the whole
language of Riemannian geometry is not very helpful in computing
the heat kernel asymptotics. That is why, until now, there are no
explicit general formulas for the coefficient $A_{1}$. A class of
so-called natural non-Laplace type operators was studied in
\cite{avramidi01,avramidi02} where this coefficient was computed
explicitly.

The main goal of this paper is to study the action of Spectral
Matrix Gravity in the weak deformation limit and to describe the
corresponding corrections to Einstein equations. For the sake of
completeness we will briefly describe, in the next section, the
properties of some geometric two-points quantities that are widely
used in the calculation of the heat kernel asymptotics.

%=================================================================
\section{Geometric Background}
\setcounter{equation}0

We introduce now some geometric two-point quantities following
mainly \cite{avramidi91,avramidi00} (see also,
\cite{synge60,dewitt03,avramidi00,avramidi91,avramidi94}). Let
$x^{\prime}$ be a fixed point on the manifold $M$. There always
exists a sufficiently small neighborhood of the point $x^{\prime}$
such that every point $x$ in this neighborhood can be connected to
the point $x^{\prime}$ by a unique geodesic. In the following,
then, we will restrict ourselves to this neighborhood. Moreover we
will denote tensor indices at point $x^{\prime}$ by primed
letters. The non-primed and primed indices will be raised and
lowered by the metric at the points $x$ and $x^{\prime}$
respectively. Also, the derivative with respect to the coordinates
of the point $x^{\prime}$ will be denoted by primed indices as
well. We will use the standard notation of square brackets to
denote the coincidence limit of two-point functions, more
precisely, for any function of $x$ and $x^{\prime}$ we define
\begin{equation}
\label{aa}
[f](x)\equiv\lim_{x\rightarrow
x^{\prime}}f(x,x^{\prime})\;.
\end{equation}

First of all, we define the world function $\sigma(x,x^{\prime})$
as one half of the square of the length of
the geodesic between the points $x$ and $x^{\prime}$.
Next, we define the first and second derivatives of the world
function
\begin{equation}
\label{4}
\sigma_{\mu}=\nabla_{\mu}\sigma,
\qquad
\sigma_{\mu^{\prime}}=\nabla_{\mu^{\prime}}\sigma\;,
\end{equation}
\begin{equation}
\label{c}
\xi^{\mu}{}_{\nu}=\nabla_{\nu}\nabla^\mu\sigma
\qquad
\eta^{\mu^{\prime}}{}_{\nu}=\nabla_{\nu}\nabla^{\mu'}\sigma\;,
\end{equation}
and the Van Vleck-Morette determinant
\begin{equation}
\label{e}
\Delta(x,x^{\prime})=e^{2\zeta(x,x')}
=g^{-\frac{1}{2}}(x)\det[-\nabla_{\mu}\nabla_{\nu^{\prime}}
\sigma(x,x^{\prime})]g^{-\frac{1}{2}}(x^{\prime})\;,
\end{equation}
where $\zeta=\log \Delta^{1/2}$.
Now, let $\varphi$ be a section of a vector bundle $V$, $\nabla_{\mu}$
be a connection on the vector bundle $V$ and
$\mathcal{R}_{\mu\nu}$ be the curvature of this connection.
%, that is,
%\begin{equation}
%\label{d}
%[\nabla_{\mu},\nabla_{\nu}]\varphi
%=\mathcal{R}_{\mu\nu}\varphi\;.
%\end{equation}
Next, let $\mathcal{P}(x,x')$ be the parallel
displacement operator
of the section $\varphi$ along the geodesic from
$x^{\prime}$ to $x$; obviously,
\begin{equation}
\label{we}
[\nabla_{\mu},\nabla_{\nu}]\mathcal{P}
=\mathcal{R}_{\mu\nu}\mathcal{P}\;.
\end{equation}
Finally, we define the first derivative of the operator of parallel
transport
\begin{equation}
\label{25}
E_{\nu}=\mathcal{P}^{-1}\nabla_\nu\mathcal{P}\;.
\end{equation}

One can show that these functions satisfy the equations
\cite{avramidi91,avramidi00}
\begin{equation}
\label{3}
\sigma=\frac{1}{2}\sigma_{\mu}\sigma^{\mu}
=\frac{1}{2}\sigma^{\mu^{\prime}}\sigma_{\mu^{\prime}}\;,
\end{equation}
\begin{equation}
\label{b}
\xi^{\mu}{}_{\nu}\sigma^{\nu}
=\sigma^{\mu}\;,\qquad\eta^{\mu^{\prime}}{}_{\nu}\sigma^{\nu}=\sigma^{\mu^{\prime}}\;,
\qquad
\eta^{\mu^{\prime}}{}_{\nu}\sigma_{\mu^{\prime}}=\sigma_{\nu}\;,
\end{equation}
\begin{equation}
\label{f}
\sigma^{\mu}\nabla_{\mu}\zeta
=\frac{1}{2}(n-\xi^\mu{}_\mu)\;.
\end{equation}
\begin{equation}
\label{9}
\sigma^\mu\nabla_\mu\mathcal{P}=\sigma^\mu E_\mu=0\;,
\end{equation}
and the boundary
conditions
\begin{equation}
\label{5}
[\sigma]=[\sigma^{\mu}]=[\sigma^{\mu^{\prime}}]=0\;,
\end{equation}
%\begin{equation}
%\label{6}
%[\nabla_{\nu}\sigma^{\mu}]=\delta^{\mu}{}_{\nu}\;\;,\;\;[\nabla_{\nu}%\sigma^{\mu^{\prime}}]
%=-\delta^{\mu}{}_{\nu}\;.
%\end{equation}
\begin{equation}
\label{ca}
[\xi^{\mu}{}_{\nu}]
=\delta^{\mu}{}_{\nu}\;,
\qquad[\eta^{\mu^{\prime}}{}_{\nu}]=-\delta^{\mu}{}_{\nu}\;.
\end{equation}
\begin{equation}
\label{g}
[\Delta]=1\;, \qquad
[\zeta]=0\,,
\end{equation}
\begin{equation}
\label{10}
[\mathcal{P}]=\mathbb{I}\;.
\end{equation}

The coincidence limits of higher derivatives of the two-point
functions introduced above are expressed in terms of the
curvature. In particular, the ones that we will need below are
\cite{avramidi00,avramidi91,dewitt03,dewitt65}
\bea
\label{27} [\zeta_{;\;\mu}]&=&0\\{}
[\zeta_{;\;\mu\nu}]&=&\frac{1}{6}R_{\mu\nu}\;,\\
\left[\eta^{\mu^{\prime}}{}_{\nu;\;\alpha}\right]&=&0\;,\\
\left[\eta^{\mu^{\prime}}{}_{\nu;\;\alpha\beta}\right]
&=&-\frac{1}{3}(R^{\mu}{}_{\nu\alpha\beta}
+R^{\mu}{}_{\alpha\nu\beta})\;,\\{}
[E_{\nu}]&=&0\\{}
[E_{\nu;\;\mu}]&=&-\frac{1}{2}\mathcal{R}_{\nu\mu}\;,
\eea
where $R^{\mu}_{\;\;\nu\rho\sigma}$ is the Riemann tensor and
$R_{\mu\nu}=R^\alpha{}_{\mu\alpha\nu}$ is the Ricci tensor.

%=================================================================
\section{Heat Kernel}
\setcounter{equation}0

We will use a method for the calculation of the heat kernel
developed in \cite{avramidi94,avramidi01}, which is based on a
covariant Fourier transform proposed in
\cite{avramidi91,avramidi00}. The heat kernel for the operator $L$
is the kernel of the heat semigroup, that is,
\begin{equation}\label{11}
U(t|x,x^{\prime})=\exp({-tL})
\mathcal{P}(x,x^{\prime})\delta(x,x^{\prime})\;,
\end{equation}
where $\delta(x,x')$ is the delta-function (in the density form).
Now, following \cite{avramidi91} we define the covariant Fourier
transform as follows. First, we define
\begin{equation}
\label{12}
\delta(x,x^{\prime})=
%g^{-\frac{1}{4}}(x)g^{-\frac{1}{4}}(x^{\prime})%
\Delta^{\frac{1}{2}}(x,x^{\prime})
\int\limits_{\mathbb{R}^{n}}\frac{d\xi}{(2\pi)^{n}}\;\exp{\{\imath
\xi_{\mu^{\prime}}\sigma^{\mu^{\prime}}(x,x^{\prime})\}}\;.
\end{equation}
Then, by using this representation and the equation
\begin{displaymath}
\exp\{-tL\}\;f=f\;\exp\{-t(f^{-1}Lf)\}\;,
\end{displaymath}
we obtain
\begin{equation}
\label{13}
U(t|x,x^{\prime})=
%g^{-\frac{1}{4}}(x)g^{-\frac{1}{4}}(x^{\prime})%
\Delta^{\frac{1}{2}}(x,x^{\prime})\mathcal{P}(x,x^{\prime})
\int\limits_{\mathbb{R}^{n}}\frac{d\xi}{(2\pi)^{n}}\exp\{{\imath
\xi_{\mu^{\prime}}\sigma^{\mu^{\prime}}(x,x^{\prime})}\}\Phi(t|k,x,x^{\prime})\;,
\end{equation}
where
\begin{equation}\label{14}
\Phi(t|k,x,x^{\prime})=\exp(-tA)\cdot\mathbb{I}\;,
\end{equation}
\begin{equation}\label{15}
A=e^{-\imath
\xi_{\mu^{\prime}}\sigma^{\mu^{\prime}}}\mathcal{P}^{-1}
\Delta^{-\frac{1}{2}}L\Delta^{\frac{1}{2}}\mathcal{P}e^{\imath
\xi_{\mu^{\prime}}\sigma^{\mu^{\prime}}}\;.
\end{equation}
By using the coincidence limits of the two-point functions
we obtain the heat kernel diagonal
\begin{equation}
\label{17}
U(t|x,x)=\int\limits_{\mathbb{R}^{n}}\frac{d\xi}{(2\pi)^{n}}\;
%g^{-\frac{1}{2}}%
\Phi(t|k,x,x)\;.
\end{equation}

Now, we represent the operator $L$ in the form
\begin{equation}
\label{18}
L=-\rho^{-1}\nabla_{\mu}\rho
a^{\mu\nu}\rho\nabla_{\nu}\rho^{-1}+{Q}\;,
\end{equation}
where $a^{\mu\nu}$ is a matrix-valued symmetric tensor of type
$(2,0)$, $\rho$ is a matrix-valued density of weight $1/2$ and
${Q}$ is a matrix-valued function. By substituting the
operator (\ref{18}) in equation (\ref{15}), we get
\begin{equation}
\label{19}
A=-e^{-\imath
\xi_{\mu^{\prime}}\sigma^{\mu^{\prime}}}
\Delta^{-\frac{1}{2}}\tilde\rho^{-1}
\mathcal{P}^{-1}\nabla_{\mu}\mathcal{P}
\tilde\rho\tilde a^{\mu\nu}\tilde \rho
\mathcal{P}^{-1}\nabla_{\nu}\mathcal{P}
\tilde\rho^{-1}\Delta^{\frac{1}{2}}e^{\imath
\xi_{\mu^{\prime}}\sigma^{\mu^{\prime}}}
+\tilde Q\;,
\end{equation}
where $\tilde a^{\mu\nu}={\cal P}^{-1}a^{\mu\nu}{\cal P}$,
$\tilde\rho={\cal P}^{-1}\rho{\cal P}$ and
$\tilde Q=\mathcal{P}^{-1}{Q}\mathcal{P}$.
We rewrite this operator in a more
convenient form
\begin{equation}\label{i}
A=-\bar{X}_{\mu}\tilde a^{\mu\nu}X_{\nu}
+\tilde Q\;,
\end{equation}
where
\begin{eqnarray}\label{l}
X_{\nu}&=&e^{-\imath
\xi_{\mu^{\prime}}\sigma^{\mu^{\prime}}}\Delta^{-\frac{1}{2}}\tilde\rho
\mathcal{P}^{-1}\nabla_{\nu}\mathcal{P}
\tilde\rho^{-1}\Delta^{\frac{1}{2}}e^{\imath
\xi_{\mu^{\prime}}\sigma^{\mu^{\prime}}}\;,\nonumber\\
\bar{X}_{\mu}&=&e^{-\imath
\xi_{\mu^{\prime}}\sigma^{\mu^{\prime}}}\Delta^{-\frac{1}{2}}
\tilde\rho^{-1}
\mathcal{P}^{-1}\nabla_{\mu}\mathcal{P}
\tilde\rho\Delta^{\frac{1}{2}}e^{\imath
\xi_{\mu^{\prime}}\sigma^{\mu^{\prime}}}\;,
\end{eqnarray}
It is useful to introduce, now, two quantities
\begin{equation}\label{23}
C_{\nu}=-\tilde\rho_{;\;\nu}\tilde\rho^{-1}
\qquad \textrm{and} \qquad
\bar{C_{\nu}}=-\tilde\rho^{-1}\tilde\rho_{;\;\nu}\;.
\end{equation}
Then we get
\begin{equation}\label{21}
X_{\nu}=\nabla_{\nu}+C_{\nu}+\zeta_{;\;\nu}+E_{\nu}+\imath
\xi_{\mu^{\prime}}\eta^{\mu^{\prime}}{}_{\nu}\;,
\end{equation}
\begin{equation}\label{21a}
\bar{X}_{\mu}=\nabla_{\mu}-\bar{C}_{\mu}+\zeta_{;\;\mu}+E_{\mu}+\imath
\xi_{\nu^{\prime}}\eta^{\nu^{\prime}}{}_{\mu}\;,
\end{equation}
and
\begin{equation}
\label{22}
A=-(\nabla_{\mu}-\bar{C}_{\mu}+\zeta_{;\;\mu}+E_{\mu}+\imath
\xi_{\rho^{\prime}}\eta^{\rho^{\prime}}{}_{\mu})
\tilde a^{\mu\nu}(\nabla_{\nu}+C_{\nu}+\zeta_{;\;\nu}+E_{\nu}+\imath
\xi_{\rho^{\prime}}\eta^{\rho^{\prime}}{}_{\nu})
+\tilde Q\;.
\end{equation}
Finally, a straightforward calculation gives
\begin{equation}
\label{29}
A=H+K+\mathcal{L}\;.
\end{equation}
Here
\bea
\label{30}
H&=&\xi_{\alpha^{\prime}}\xi_{\beta^{\prime}}\eta^{\alpha^{\prime}}{}_{\mu}
\eta^{\beta^{\prime}}{}_{\nu}
\tilde a^{\mu\nu}\;,\\
\label{61a}
K&=&-\imath\xi_{\rho^{\prime}}(B^{\rho^{\prime}\nu}\nabla_{\nu}+G^{\rho^{\prime}})\;,\\
\label{60}
\mathcal{L}&=&-\bar{\mathcal{D}}_{\mu}
\tilde a^{\mu\nu}\mathcal{D}_{\nu}+\tilde Q\;,
\eea
where
\begin{eqnarray}
\label{61b}
B^{\rho^{\prime}\nu}&=&2\eta^{\rho^{\prime}}{}_{\mu}
\tilde a^{\mu\nu}\;,\nonumber\\
G^{\rho^{\prime}}&=&
\tilde a^{\mu\nu}{}_{;\;\mu}\eta^{\rho^{\prime}}{}_{\nu}
+\tilde a^{\mu\nu}\eta^{\rho^{\prime}}{}_{\nu;\;\mu}
-\bar{C}_{\mu}\tilde a^{\mu\nu}\eta^{\rho^{\prime}}{}_{\nu}
+\eta^{\rho^{\prime}}{}_{\nu}\tilde a^{\mu\nu}C_{\mu}\nonumber\\
&+&
E_{\mu}\tilde a^{\mu\nu}\eta^{\rho^{\prime}}{}_{\nu}
+\eta^{\rho^{\prime}}{}_{\nu}\tilde a^{\mu\nu}E_{\mu}
+2\zeta_{;\;\mu}\tilde a^{\mu\nu}\eta^{\rho^{\prime}}{}_{\nu}\;,\\
\label{61}
\bar{\mathcal{D}}_{\mu}&=&\nabla_{\mu}+\bar{\mathcal{A}}_{\mu}=\nabla_{\mu}-\bar{C}_{\mu}+\zeta_{;\;\mu}+E_{\mu}\;,\nonumber\\
\mathcal{D}_{\nu}&=&\nabla_{\nu}+\mathcal{A}_{\nu}=\nabla_{\nu}+C_{\nu}+\zeta_{;\;\nu}+E_{\nu}\;,
\end{eqnarray}
with $C_{\nu}$, $\bar{C}_{\mu}$, $\zeta$ and $E_{\mu}$ defined in
(\ref{23}), (\ref{e}) and (\ref{25}).

More explicitly we can also
write that
\begin{equation}\label{30a}
\mathcal{L}=
-\tilde a^{\mu\nu}\nabla_{\mu}\nabla_{\nu}+\mathcal{Y}^{\mu}\nabla_{\mu}+\mathcal{Z}\;,
\end{equation}
where
\bea
\label{30b}
\mathcal{Y}^{\mu}
&=&
-\tilde a^{\mu\nu}{}_{;\;\nu}
+\bar{C}_{\nu}\tilde a^{\mu\nu}
-\tilde a^{\mu\nu}C_{\nu}
-2\tilde a^{\mu\nu}\zeta_{;\;\nu}
-\tilde a^{\mu\nu}E_{\nu}
-E_{\nu}\tilde a^{\mu\nu}\;,\\
\label{32}
\mathcal{Z}&=&
-\tilde a^{\mu\nu}{}_{;\;\mu}C_{\nu}
-\tilde a^{\mu\nu}C_{\nu;\;\mu}
+\bar{C}_{\mu}\tilde a^{\mu\nu}C_{\nu}
-\tilde a^{\mu\nu}{}_{;\;\mu}\zeta_{;\;\nu}
-\tilde a^{\mu\nu}\zeta_{;\;\mu\nu}
+\bar{C}_{\mu}\tilde a^{\mu\nu}\zeta_{;\;\nu}\nonumber\\
&-&\zeta_{;\;\mu}\tilde a^{\mu\nu}C_{\nu}
-\zeta_{;\;\mu}\tilde a^{\mu\nu}\zeta_{;\;\nu}
-\tilde a^{\mu\nu}{}_{;\;\mu}E_{\nu}
-\tilde a^{\mu\nu}E_{\nu;\;\mu}
+\bar{C}_{\mu}\tilde a^{\mu\nu}E_{\nu}
-E_{\mu}\tilde a^{\mu\nu}C_{\nu}\nonumber\\
&-&E_{\mu}\tilde a^{\mu\nu}E_{\nu}
-\zeta_{;\;\nu}\tilde a^{\mu\nu}E_{\mu}
-\zeta_{;\;\nu}E_{\mu}\tilde a^{\mu\nu}+\tilde Q\;.
\eea

Thus, by using the eq. (\ref{29}) we obtain
\begin{equation}
\label{33}
U(t|x,x)=\int\limits_{\RR^n}\frac{d\xi}{(2\pi)^{n}}\;%g^{-\frac{1}{2}}%
e^{-t(H+K+\mathcal{L})}\cdot\mathbb{I}\Big|_{x=x^{\prime}}\;,
\end{equation}
which, by scaling the integration variable $\xi\rightarrow
t^{-\frac{1}{2}}\xi$, takes the form
\begin{equation}\label{34}
U(t|x,x)=(4\pi
t)^{-\frac{n}{2}}\int\limits_{\mathbb{R}^{n}}\frac{d\xi}{\pi^{\frac{n}{2}}}\;%g^{-\frac{1}{2}}%
\exp(-H-\sqrt{t}K-t\mathcal{L})\cdot\mathbb{I}\Big|_{x=x^{\prime}}\;.
\end{equation}
It is convenient, to rewrite this equation as
\begin{equation}\label{34a}
U(t|x,x)=(4\pi
t)^{-\frac{n}{2}}\int\limits_{\mathbb{R}^{n}}\frac{d\xi}{\pi^{\frac{n}{2}}}\;%g^{-\frac{1}{2}}%
\;e^{-|\xi|^{2}}\exp(-\tilde{H}-\sqrt{t}K-t\mathcal{L})\cdot\mathbb{I}\Big|_{x=x^{\prime}}\;,
\end{equation}
where $|\xi|^{2}=g^{\mu\nu}\xi_{\mu}\xi_{\nu}$ and
\begin{equation}\label{34b}
\tilde{H}=H-|\xi|^{2}\;.
\end{equation}

In order to evaluate the first three coefficients of the
asymptotic expansion of (\ref{34a}) as $t\rightarrow 0$ we use the
Volterra series for the exponent of a sum of two non-commuting operators
\begin{equation}
\label{35}
e^{A+B}=e^{A}+\sum_{k=1}^{\infty}\int\limits_{0}^{1}d\tau_{k}\int\limits_{0}^{\tau_{k}}d\tau_{k-1}\cdots\int\limits_{0}^{\tau_{2}}d\tau_{1}e^{(1-\tau_{k})A}Be^{(\tau_{k}-\tau_{k-1})A}\cdots
 e^{(\tau_{2}-\tau_{1})A}Be^{\tau_{1}A}\;.
\end{equation}
We obtain
\begin{equation}\label{35b}
\exp(-\tilde{H}-\sqrt{t}K-t\mathcal{L})=e^{-\tilde{H}}-\sqrt{t}\;\Omega+t\Psi+O(t^{\frac{3}{2}})\;,
\end{equation}
where
\begin{equation}\label{35c}
\Omega=\int\limits_{0}^{1}d\tau_{1}e^{-(1-\tau_{1})\tilde{H}}Ke^{-\tau_{1}\tilde{H}}\;,
\end{equation}
and
\begin{equation}\label{35d}
\Psi=\int\limits_{0}^{1}d\tau_{2}\int\limits_{0}^{\tau_{2}}d\tau_{1}e^{-(1-\tau_{2})\tilde{H}}Ke^{-(\tau_{2}-\tau_{1})\tilde{H}}Ke^{-\tau_{1}\tilde{H}}-\int\limits_{0}^{1}d\tau_{1}e^{-(1-\tau_{1})\tilde{H}}\mathcal{L}e^{-\tau_{1}\tilde{H}}\;.
\end{equation}

We are only interested in the terms $a_{0}$ and $a_{1}$ of the
heat kernel expansion, namely the terms of zero order and linear
in the parameter $t$. These terms can be written, respectively, as
\begin{eqnarray}\label{35e}
a_{0}&=&g^{\frac{1}{2}}\tilde{a}_{0}\;,\\
a_{1}&=&g^{\frac{1}{2}}\tilde{a}_{1}\;,
\end{eqnarray}
where
\begin{eqnarray}\label{36}
\tilde{a}_{0}&=&\int\limits_{\mathbb{R}^{n}}\frac{d\xi}{\pi^{\frac{n}{2}}}\;g^{-\frac{1}{2}}\;e^{-|\xi|^{2}}\exp\{-\tilde{H}\}\cdot\mathbb{I}\Big|_{x=x^{\prime}}\;,\\
\label{360}
\tilde{a}_{1}&=&\int\limits_{\mathbb{R}^{n}}\frac{d\xi}{\pi^{\frac{n}{2}}}\;g^{-\frac{1}{2}}\;e^{-|\xi|^{2}}\Psi\cdot\mathbb{I}\Big|_{x=x^{\prime}}\;.
\end{eqnarray}

The term $t^{\frac{1}{2}}$ of the heat kernel expansion vanishes
identically. This happens because the heat kernel coefficients are
defined as $\xi$-integrals over the whole $\mathbb{R}^{n}$ and the
term $\Omega$ in (\ref{35c}) is an odd function of $\xi$.

%=================================================================
\section{Evaluation of the Heat Kernel Coefficients}
\setcounter{equation}0

\subsection{Local Coefficient $\tilde{a}_{0}$}

We will evaluate the heat kernel coefficients $A_{0}$
and $A_{1}$ using the perturbation theory.
The main idea is to introduce a small deformation parameter
$\lambda$ and evaluate the non-commutative corrections to the
action of Spectral matrix Gravity. For this purpose we write the
matrix $a_{\mu\nu}$ as
\begin{equation}\label{38}
a^{\mu\nu}=g^{\mu\nu}\,\mathbb{I}+\lambda h^{\mu\nu}\;,
\end{equation}
where $h^{\mu\nu}$ is a traceless
matrix-valued tensor field (a non-commutative
perturbation of the Riemannian metric), satisfying
\begin{equation}\label{38a}
\textrm{tr}_{V} h^{\mu\nu}=0\;.
\end{equation}
Furthermore, we parametrize the matrix-valued density $\rho$
introduced in (\ref{18}) as
\begin{equation}\label{39}
\rho=g^{\frac{1}{4}}e^{\phi}e^{\lambda\sigma}\;.
\end{equation}
Here $\sigma$ is a
traceless matrix-valued scalar field and $\phi$ is a scalar field.
({\it Do not confuse it with the world function introduced in the previous
sections!})
Finally, we also decompose the
endomorphism $Q$,
\begin{equation}\label{re}
Q=q\cdot\mathbb{I}+\lambda \Theta\;,
\end{equation}
where $\Theta$ is a traceless matrix-valued scalar field.

Now we expand all the quantities in powers of $\lambda$.
On doing so the matrix $\rho$ and its
inverse read
\begin{eqnarray}\label{40}
\rho&=&g^{\frac{1}{4}}\;e^{\phi}\left(1+\lambda\sigma+\frac{\lambda^{2}}{2}\sigma^{2}\right)+O(\lambda^{3})\;,\nonumber\\
\rho^{-1}&=&g^{-\frac{1}{4}}\;e^{-\phi}\left(1-\lambda\sigma+\frac{\lambda^{2}}{2}\sigma^{2}\right)+O(\lambda^{3})\;,
\end{eqnarray}
and its derivative is
\begin{equation}\label{41}
g^{-\frac{1}{4}}\rho_{;\;\nu}=e^{\phi}\left[\lambda\sigma_{;\;\nu}+\frac{\lambda^{2}}{2}(\sigma_{;\;\nu}\sigma+\sigma\sigma_{;\;\nu})\right]+e^{\phi}\phi_{;\;\nu}\left(1+\lambda\sigma+\frac{\lambda^{2}}{2}\sigma^{2}\right)+O(\lambda^{3})\;.
\end{equation}
{}From the last two expressions one can easily evaluate the
operators $C_{\nu}$ and $\bar{C}_{\nu}$ obtaining explicitly
\begin{eqnarray}\label{42}
C_{\nu}=-\tilde\rho_{;\;\nu}\tilde\rho^{-1}
=-\phi_{;\;\nu}
-\lambda\tilde\sigma_{;\;\nu}
+\frac{\lambda^{2}}{2}[\tilde\sigma_{;\;\nu},\tilde\sigma]
+O(\lambda^{3})\;,
\end{eqnarray}
\begin{eqnarray}\label{43}
\bar{C}_{\nu}=-\tilde\rho^{-1}\tilde\rho_{;\;\nu}
=-\phi_{;\;\nu}-\lambda\tilde\sigma_{;\;\nu}
-\frac{\lambda^{2}}{2}[\tilde\sigma_{;\;\nu},\tilde\sigma]
+O(\lambda^{3})\;,
\end{eqnarray}
where $\tilde \sigma={\cal P}^{-1}\sigma{\cal P}$.

The operators $\tilde{H}$, $K$ and $\mathcal{L}$ introduced above
in (\ref{30}), (\ref{61a}) and (\ref{60}) depend on the
deformation parameter $\lambda$ as well. By expanding them in
terms of the deformation parameter we get
\begin{eqnarray}\label{44}
\tilde{H}&=&H_{0}+\lambda H_{1}\;,\nonumber\\
K&=&K_{0}+\lambda K_{1}+\lambda^{2}K_{2}+O(\lambda^{2})\;,\nonumber\\
\mathcal{L}&=&\mathcal{L}_{0}+\lambda
\mathcal{L}_{1}+\lambda^{2}\mathcal{L}_{2}+O(\lambda^{2})\;,
\end{eqnarray}
where (after defining $\tilde h^{\mu\nu}={\cal P}^{-1}h^{\mu\nu}{\cal P}$ and
$\tilde \Theta={\cal P}^{-1}\Theta{\cal P}$)
\begin{eqnarray}\label{45}
H_{0}&=&\xi_{\alpha^{\prime}}\xi_{\beta^{\prime}}
(\eta^{\alpha^{\prime}}{}_{\mu}\eta^{\beta^{\prime}}{}_{\nu}g^{\mu\nu}
-g^{\alpha^{\prime}\beta^{\prime}})\;,\\[3pt]
\label{450}
H_{1}&=&\xi_{\alpha^{\prime}}\xi_{\beta^{\prime}}\eta^{\alpha^{\prime}}{}_{\mu}
\eta^{\beta^{\prime}}{}_{\nu}
\tilde h^{\mu\nu}\;,\\[12pt]
K_{0}&=&-\imath\xi_{\alpha^{\prime}}(2\eta^{\alpha^{\prime}}{}_{\mu}g^{\mu\nu}\nabla_{\nu}
+2\eta^{\alpha^{\prime}}{}_{\mu}g^{\mu\nu}\zeta_{;\;\nu}
+\eta^{\alpha^{\prime}}{}_{\;\;\mu}{}^{;\;\mu}
+E^{\mu}\eta^{\alpha^{\prime}}{}_{\mu}
+\eta^{\alpha^{\prime}}{}_{\mu}E^{\mu})\;,\\[3pt]
K_{1}&=&
-\imath\xi_{\alpha^{\prime}}(2\eta^{\alpha^{\prime}}{}_{\mu}
\tilde h^{\mu\nu}\nabla_{\nu}
+\eta^{\alpha^{\prime}}{}_{\mu}
\tilde h^{\mu\nu}{}_{;\nu}
+2\eta^{\alpha^{\prime}}{}_{\mu}
\tilde h^{\mu\nu}\zeta_{;\;\nu}
+h^{\mu\nu}\eta^{\alpha^{\prime}}{}_{\mu;\;\nu}
+\eta^{\alpha^{\prime}}{}_{\mu}
\tilde h^{\mu\nu}E_{\nu}\nonumber\\
&+&
\eta^{\alpha^{\prime}}{}_{\mu}E_{\nu}
\tilde h^{\mu\nu})\;,\\[3pt]
K_{2}&=&-\imath\xi_{\alpha^{\prime}}[\eta^{\alpha^{\prime}}_{\;\;\mu}(
[\tilde\sigma_{;\;\nu}, \tilde h^{\mu\nu}]+[\tilde\sigma^{;\;\mu},\tilde\sigma])]\;,
\end{eqnarray}
%And finally for the operator $\mathcal{L}$ we have
\begin{eqnarray}\label{50}
\mathcal{L}_{0}&=&-\nabla^{2}
-(2\zeta^{;\;\mu}
+2E^{\mu})\nabla_{\mu}
+\phi_{;\;\mu}{}^{;\;\mu}
+\phi_{;\;\mu}\phi^{;\;\mu}
-\zeta_{;\;\mu}{}^{;\;\mu}\nonumber\\
&-&\zeta^{;\;\mu}\zeta_{;\;\mu}-E_{\mu}{}^{;\;\mu}
-E^{\mu}E_{\mu}
-2\zeta_{;\;\mu}E^{\mu}+q\;,\\[3pt]
\mathcal{L}_{1}
&=&-\tilde h^{\mu\nu}\nabla_{\mu}\nabla_{\nu}
-[\tilde h^{\mu\nu}{}_{;\;\nu}
+2\zeta_{;\;\nu}\tilde h^{\mu\nu}
+(\tilde h^{\mu\nu}E_{\nu}
+E_{\nu}\tilde h^{\mu\nu})]\nabla_{\mu}
+\tilde \sigma_{;\;\mu}{}^{;\;\mu}\nonumber\\
&+&\tilde h^{\mu\nu}{}_{;\;\mu}\phi_{;\;\nu}
+\tilde h^{\mu\nu}\phi_{;\;\mu\nu}
+2\phi^{;\;\nu}\tilde\sigma_{;\;\nu}
+\phi_{;\;\mu}\tilde h^{\mu\nu}\phi_{;\;\nu}
-\tilde h^{\mu\nu}{}_{;\;\mu}\zeta_{;\;\nu}
\nonumber\\
&-&\tilde h^{\mu\nu}\zeta_{;\;\mu\nu}
-\zeta_{;\;\mu}\tilde h^{\mu\nu}\zeta_{;\;\nu}
-\tilde h^{\mu\nu}{}_{;\;\mu}E_{\nu}
-\tilde h^{\mu\nu}E_{\nu;\;\mu}
+[E_{\mu},\tilde \sigma^{;\;\mu}]\nonumber\\
&+&\phi_{;\;\mu}[E_{\nu},\tilde h^{\mu\nu}]
-E_{\mu}\tilde h^{\mu\nu}E_{\nu}
-E_{\mu}\tilde h^{\mu\nu}\zeta_{;\;\nu}
-\zeta_{;\;\mu}\tilde h^{\mu\nu}E_{\nu}
+\tilde\Theta\;,\\[3pt]
\mathcal{L}_{2}
&=&([\tilde h^{\mu\nu},\tilde\sigma_{;\;\nu}]
-[\tilde\sigma^{;\;\mu},\tilde\sigma])\nabla_{\mu}
-\frac{1}{2}[\tilde\sigma_{;\;\mu}{}^{;\;\mu},\tilde\sigma]
+\tilde h^{\mu\nu}{}_{;\;\mu}\tilde\sigma_{;\;\nu}
+\tilde h^{\mu\nu}\tilde\sigma_{;\;\mu\nu}
\nonumber\\
&+&\tilde \sigma^{;\;\nu}\tilde\sigma_{;\;\nu}
+\tilde \sigma_{;\;\nu}\tilde h^{\mu\nu}\phi_{;\;\mu}
+\phi_{;\;\nu}\tilde h^{\mu\nu}\tilde \sigma_{;\;\mu}
-\frac{1}{2}[\tilde \sigma^{;\;\mu},\tilde \sigma]\zeta_{;\;\mu}
-\frac{1}{2}\zeta_{;\;\mu}[\tilde \sigma^{;\;\mu},\tilde \sigma]
\nonumber\\
&+&\zeta_{;\;\mu}\tilde h^{\mu\nu}\tilde \sigma_{;\;\nu}
-\tilde \sigma_{;\;\nu}\tilde h^{\mu\nu}\zeta_{;\;\mu}
-\frac{1}{2}([\tilde \sigma_{;\;\mu},\tilde \sigma]E^{\mu}
+E^{\mu}[\tilde \sigma_{;\;\mu},\tilde \sigma])\nonumber\\
&+&E_{\mu}\tilde h^{\mu\nu}\tilde \sigma_{;\;\nu}
-\tilde \sigma_{;\;\mu}\tilde h^{\mu\nu}E_{\nu}\;.
\label{500}
\end{eqnarray}

In the framework of perturbation theory we write, then, the
coefficients $\tilde{a}_{0}$ and $\tilde{a}_{1}$ of the heat
kernel expansion in (\ref{36}) and (\ref{360}) in terms of the
deformation parameter $\lambda$, namely
\begin{eqnarray}\label{53}
\tilde{a}_{0}(\lambda)&=&a_{0}^{(0)}+\lambda
a_{0}^{(1)}+\lambda^{2}a_{0}^{(2)}+O(\lambda^{3})\nonumber\\
\tilde{a}_{1}(\lambda)&=&a_{1}^{(0)}+\lambda
a_{1}^{(1)}+\lambda^{2}a_{1}^{(2)}+O(\lambda^{3})\;.
\end{eqnarray}
By using the explicit formulas obtained in (\ref{45}) through
(\ref{500}), we will be able to evaluate all the coefficients of
the Taylor expansions in (\ref{53}).

Next, we introduce a notation that will be useful in the following
calculations. Let $f$ be a function of $\xi$. We define the
Gaussian average of the function $f$ as
\begin{equation}\label{54}
\bra{f}\ket{}=\int\limits_{\mathbb{R}^{n}}\frac{d\xi}{\pi^{\frac{n}{2}}}\;g^{-\frac{1}{2}}\;e^{-|\xi|^{2}}f(\xi)\;.
\end{equation}
The Gaussian averages of the polynomials is well known
\begin{eqnarray}\label{55}
\bra{\xi_{\mu_{1}}\cdots\xi_{\mu_{2n+1}}}\ket{}&=&0\;,\nonumber\\
\bra{\xi_{\mu_{1}}\cdots\xi_{\mu_{2n}}}\ket{}&=&\frac{(2n)!}{2^{2n}n!}g_{(\mu_{1}\mu_{2}}\cdots
g_{\mu_{2n-1}\mu_{2n})}\;,
\end{eqnarray}
where the parentheses $(\;)$ denote the symmetrization over all the
included indices.

For the coefficient of order zero of the heat kernel expansion we
consider the first equation in (\ref{53}). From the formula
(\ref{36}), it is easy to see that the only non-vanishing
contribution to $\tilde{a}_{0}$ is
\begin{equation}\label{56}
\tilde{a}_{0}=\left<\left(1-\lambda H_1
+\frac{\lambda^{2}}{2}H_1^{2}\right)
\right>\Bigg|_{x=x^{\prime}}+O(\lambda^{3})\;.
\end{equation}
By using the equations (\ref{45}), (\ref{450}) and after taking
the coincidence limit we obtain the expression
\begin{equation}\label{57}
\tilde{a}_{0}=\left<\left(\mathbb{I}-\lambda
h^{\mu\nu}\xi_{\mu}\xi_{\nu}+\frac{1}{2}\lambda^{2}h^{\mu\nu}h^{\rho\sigma}\xi_{\mu}\xi_{\nu}\xi_{\rho}\xi_{\sigma}\right)\right>+O(\lambda^{3})\;,
\end{equation}
and then, by performing the Gaussian averages, we get
\begin{equation}\label{58}
\tilde{a}_{0}=1-\frac{\lambda}{2}h+\frac{\lambda^{2}}{8}(h^{2}+2h^{\mu\nu}h_{\mu\nu})+O(\lambda^{3})\;,
\end{equation}
where $h=g_{\mu\nu}h^{\mu\nu}$. In order to evaluate the global
coefficient $A_{0}$, given in (\ref{vd1}), we need the trace of
(\ref{58}). Since $h^{\mu\nu}$ is traceless we immediately obtain
\begin{equation}\label{59}
\textrm{tr}_{V}\tilde{a}_{0}
=\textrm{tr}_{V}\left(1+\frac{\lambda^{2}}{8}h^{2}+\frac{\lambda^{2}}{4}h^{\mu\nu}h_{\mu\nu}\right)+O(\lambda^{3})\;.
\end{equation}

\subsection{Local Coefficient $\tilde{a}_{1}$}

Now we evaluate the coefficient $\tilde{a}_{1}$. By using the
expressions (\ref{360}), (\ref{35d}) and (\ref{60}) we have
\begin{eqnarray}\label{62}
\textrm{tr}_{V} \tilde{a}_{1}
&=&\Biggl\{\left<\textrm{tr}_{V}\;[-e^{-\tilde{H}}Q]\right>
+\left<\textrm{tr}_{V}\;\Bigg[\int\limits_{0}^{1}d\tau_{2}\int\limits_{0}^{\tau_{2}}d\tau_{1}
e^{-(1-\tau_{2})\tilde{H}}
Ke^{-(\tau_{2}-\tau_{1})\tilde{H}}Ke^{-\tau_{1}\tilde{H}}\Bigg]\right>\nonumber\\
&-&
\left<\textrm{tr}_{V}\;\Bigg[\int\limits_{0}^{1}d\tau_{1}e^{-(1-\tau_{1})
\tilde{H}}\bar{\mathcal{D}}_{\mu}a^{\mu\nu}\mathcal{D}_{\nu}
e^{-\tau_{1}\tilde{H}}\Bigg]\right>
\Biggr\}\Bigg|_{x=x^{\prime}}\;,
\end{eqnarray}
where the first term of the expression (\ref{62}) has been
obtained by simply using the cyclic property of the trace.

In the following we will evaluate the terms in (\ref{62})
separately. We start with the simplest of them, namely the one
involving the endomorphism $Q$. By using the Taylor expansion in
$\lambda$ of $\tilde{H}$ in (\ref{44}) and the coincidence limits
(\ref{ca}) we obtain
\begin{equation}\label{63}
\left<\textrm{tr}_{V}\;[-e^{-\tilde{H}}Q]\right>\Big|_{x=x^{\prime}}=\left<\textrm{tr}_{V}\;\left(-Q+\lambda\xi_{\mu}\xi_{\nu}h^{\mu\nu}Q-\frac{\lambda^{2}}{2}\xi_{\mu}\xi_{\nu}\xi_{\rho}\xi_{\sigma}h^{\mu\nu}h^{\rho\sigma}Q\right)\right>\Bigg|_{x=x^{\prime}}+O(\lambda^{3})\;.
\end{equation}
By expanding $Q$ as in (\ref{re}) and by performing the Gaussian
averages we obtain
\begin{equation}\label{64}
\left<\textrm{tr}_V\;[-e^{-\tilde{H}}Q]\right>\Big|_{x=x^{\prime}}=-Nq+\lambda^{2}\textrm{tr}_{V}\left(\frac{1}{2}h\Theta-\frac{1}{8}h^{2}q-\frac{1}{4}h^{\mu\nu}h_{\mu\nu}q\right)+O(\lambda^{3})\;,
\end{equation}
where we used the property (\ref{38a}).

For the second term in equation (\ref{62}) we get, by using the
definition (\ref{61a}),
\begin{eqnarray}\label{65}
\lefteqn{\left<\textrm{tr}_{V}\;\Bigg[\int\limits_{0}^{1}d\tau_{2}\int\limits_{0}^{\tau_{2}}d\tau_{1}
e^{-(1-\tau_{2})\tilde{H}}Ke^{-(\tau_{2}-\tau_{1})\tilde{H}}
Ke^{-\tau_{1}\tilde{H}}\Bigg]\right>\Bigg|_{x=x^{\prime}}}
\nonumber\\
&=&-\Bigg<\textrm{tr}_{V}\;\Bigg[\xi_{\rho^{\prime}}\xi_{\sigma^{\prime}}\int\limits_{0}^{1}d\tau_{2}\int\limits_{0}^{\tau_{2}}
d\tau_{1}e^{-(1-\tau_{2})\tilde{H}}\Big(B^{\rho^{\prime}\nu}\nabla_{\nu}
e^{-(\tau_{2}-\tau_{1})\tilde{H}}B^{\sigma^{\prime}\mu}\nabla_{\mu}e^{-\tau_{1}\tilde{H}}
\nonumber\\
&&+B^{\rho^{\prime}\nu}\nabla_{\nu}e^{-(\tau_{2}-\tau_{1})\tilde{H}}
G^{\sigma^{\prime}}e^{-\tau_{1}\tilde{H}}
+G^{\rho^{\prime}}e^{-(\tau_{2}-\tau_{1})\tilde{H}}
B^{\sigma^{\prime}\mu}\nabla_{\mu}e^{-\tau_{1}\tilde{H}}
\nonumber\\
&&+G^{\rho^{\prime}}e^{-(\tau_{2}-\tau_{1})\tilde{H}}
G^{\sigma^{\prime}}e^{-\tau_{1}\tilde{H}}\Big)\Bigg]\Bigg>\Bigg|_{x=x^{\prime}}\;.
\end{eqnarray}
It is straightforward to notice that in the last expression we
need to compute first and second derivatives of the exponentials
containing the operator $\tilde{H}$. These derivatives are
computed by using integral representations, i.e. for the first
derivative we have \cite{avramidi04b}
\begin{equation}\label{66}
\nabla_{\mu}e^{-\tau
\tilde{H}}=-\beta_{\mu}(\tau)e^{-\tau\tilde{H}}\;,
\end{equation}
where
\begin{equation}\label{67}
\beta_{\mu}(\tau)=\int\limits_{0}^{\tau}ds\;e^{-s\tilde{H}}\tilde{H}_{;\;\mu}e^{s\tilde{H}}\;.
\end{equation}
This last integral can be evaluated by referring to the following
formula and by integrating over $s$ \cite{avramidi04b}
\begin{equation}\label{67b}
e^{-\tilde{H}}\tilde{H}_{;\;\mu}e^{\tilde{H}}=\sum_{k=0}^{\infty}\frac{(-1)^{k}}{k!}\underbrace{[\tilde{H},\cdots[\tilde{H}}_{k},\tilde{H}_{;\;\mu}]\cdots]\;.
\end{equation}
By expanding (\ref{67}) in $s$, up to the second order in
$\lambda$, we obtain
\begin{equation}\label{67a}
\nabla_{\mu}e^{-\tau
\tilde{H}}=-\left(\tau\tilde{H}_{;\;\mu}+\frac{1}{2}\tau^{2}[\tilde{H}_{;\;\mu},\tilde{H}]\right)e^{-\tau\tilde{H}}+O(\lambda^{3})\;.
\end{equation}
This last expression can be obtained by recalling that the
coincidence limit for $\tilde{H}$ and its derivatives is of order
$\lambda$ without the zeroth order term (see Appendix A).

For the second derivative we write \cite{avramidi04b}
\begin{eqnarray}\label{68}
\nabla_{\mu}\nabla_{\nu}e^{-\tau\tilde{H}}&=&-\int\limits_{0}^{\tau}ds_{1}e^{-(\tau-s_{1})\tilde{H}}\tilde{H}_{;\;\mu\nu}e^{-s_{1}\tilde{H}}+\nonumber\\
&+&\int\limits_{0}^{\tau}ds_{2}\int\limits_{0}^{s_{2}}ds_{1}\Big(e^{-(s_{2}-s_{1})\tilde{H}}\tilde{H}_{;\;\nu}e^{-s_{1}\tilde{H}}\tilde{H}_{;\;\mu}e^{-(\tau-s_{2})\tilde{H}}+\nonumber\\
&+&e^{-(\tau-s_{2})\tilde{H}}\tilde{H}_{;\;\mu}e^{-(s_{2}-s_{1})\tilde{H}}\tilde{H}_{;\;\nu}e^{-s_{1}\tilde{H}}\Big)\;.
\end{eqnarray}
We can express this formula in the same form as (\ref{67a}),
i.e.
\begin{equation}\label{69}
\nabla_{\mu}\nabla_{\nu}e^{-\tau\tilde{H}}=-\left(\tau\tilde{H}_{;\;\mu\nu}+\frac{1}{2}[\tilde{H}_{;\;\mu\nu},\tilde{H}]-\frac{1}{2}\tau^{2}\{\tilde{H}_{;\;\nu},\tilde{H}_{;\;\mu}\}\right)e^{-\tau\tilde{H}}+O(\lambda^{3})\;.
\end{equation}

Now that we have the expressions (\ref{66}) through (\ref{69}), we
can substitute them in (\ref{65}) and we can expand the remaining
exponentials in $\tau$ up to orders $\lambda^{2}$. After the
expansion of the exponentials and after taking the coincidence
limit, we have to evaluate the double integrals of polynomials in
$\tau_{1}$ and $\tau_{2}$ which will yield the numerical
coefficients for the various terms in (\ref{65}). The most general
double integral that we need to evaluate is the following
\begin{equation}\label{70}
I_{1}(\alpha,\beta,\gamma,\delta)=\int\limits_{0}^{1}d\tau_{2}\int\limits_{0}^{\tau_{2}}d\tau_{1}\tau_{1}^{\alpha}(1-\tau_{2})^{\beta}(\tau_{2}-\tau_{1})^{\gamma}(1-\tau_{2}+\tau_{1})^{\delta}\;,
\end{equation}
with $(\alpha,\beta,\gamma,\delta)$ being positive integers so
that the integral is well defined. The solution in closed form of
(\ref{70}) is (see Appendix B)
\begin{equation}\label{71}
I_{1}(\alpha,\beta,\gamma,\delta)=\frac{\Gamma(1+\alpha)\Gamma(1+\beta)\Gamma(1+\gamma)\Gamma(2+\alpha+\beta+\delta)}{\Gamma(3+\alpha+\beta+\gamma+\delta)\Gamma(2+\alpha+\beta)}\;,
\end{equation}
where $\Gamma(x)$ is the Euler gamma function.

By using the technical details described above and the coincidence
limits for the various terms in (\ref{65}) (see Appendix A), we
obtain
\begin{eqnarray}\label{72}
\left<\textrm{tr}_{V}\;\Bigg[\int\limits_{0}^{1}d\tau_{2}\int\limits_{0}^{\tau_{2}}d\tau_{1}
e^{-(1-\tau_{2})\tilde{H}}Ke^{-(\tau_{2}-\tau_{1})\tilde{H}}
Ke^{-\tau_{1}\tilde{H}}\Bigg]\right>\Bigg|_{x=x^{\prime}}
\nonumber\\
=\textrm{tr}_{V}\left(\Omega_{0}\right)+\lambda\textrm{tr}_{V}\left(\Omega_{1}\right)
+\lambda^{2}\textrm{tr}_{V}\left(\Omega_{2}\right)+O(\lambda^{3})\;,
\end{eqnarray}
where
\begin{eqnarray}\label{73}
\Omega_{0}&=&\frac{1}{6}R\;,\\[3pt]
\label{730}
\Omega_{1}&=&0
% this terms are traceless
%\frac{1}{6}h^{;\alpha}{}_{;\alpha}
%-\frac{1}{6}h^{\mu\nu}{}_{;\;\mu\nu}
%-\frac{1}{12}hR
%+\frac{1}{6}h^{\mu\nu}R_{\mu\nu}
\;,\\[3pt]
\label{731}
\Omega_{2}&=&-\frac{1}{16}h_{;\;\mu}h^{;\;\mu}+\frac{1}{6}h^{\mu\nu}{}_{;\;\mu}h_{;\;\nu}-\frac{1}{8}h^{\mu\nu}{}_{;\;\rho}h_{\mu\nu}{}^{;\;\rho}-\frac{1}{12}h
h^{;\alpha}{}_{;\alpha}
+\frac{1}{12}hh^{\mu\nu}{}_{;\;\mu\nu}\nonumber\\
&+&\frac{1}{6}h^{\mu\nu}h_{;\;\mu\nu}
-\frac{1}{6}h^{\mu\nu}h_{\mu\nu}{}^{;\alpha}{}_{;\alpha}
+\frac{1}{12}h^{\mu\nu;\;\rho}h_{\mu\rho;\;\nu}-\frac{1}{6}h^{\mu}{}_{\rho}h^{\nu\rho}R_{\mu\nu}+\frac{1}{48}h^{2}R\nonumber\\
&-&\frac{1}{12}hh^{\mu\nu}R_{\mu\nu}+\frac{1}{24}h^{\mu\nu}h_{\mu\nu}R\;.
\end{eqnarray}

We can finally evaluate the last term in equation (\ref{62}). By
using the definitions (\ref{60}) and (\ref{61}) we can write that
\begin{eqnarray}\label{76}
\lefteqn{\left<\textrm{tr}_{V}\;\Bigg[\int\limits_{0}^{1}d\tau_{1}e^{-(1-\tau_{1})\tilde{H}}\bar{\mathcal{D}}_{\mu}a^{\mu\nu}\mathcal{D}_{\nu}e^{-\tau_{1}\tilde{H}}\Bigg]\right>\Bigg|_{x=x^{\prime}}}\nonumber\\
&&=\Bigg<\textrm{tr}_{V}\;\Bigg[\int\limits_{0}^{1}d\tau_{1}e^{-(1-\tau_{1})\tilde{H}}\Big(a^{\mu\nu}{}_{;\;\mu}\nabla_{\nu}+a^{\mu\nu}\nabla_{\mu}\nabla_{\nu}+a^{\mu\nu}{}_{;\;\mu}\mathcal{A}_{\nu}\nonumber\\
&&+a^{\mu\nu}\mathcal{A}_{\nu;\;\mu}+a^{\mu\nu}\mathcal{A}_{\nu}\nabla_{\mu}+\bar{\mathcal{A}}_{\mu}a^{\mu\nu}\nabla_{\nu}+\bar{\mathcal{A}}_{\mu}a^{\mu\nu}\mathcal{A}_{\nu}\Big)\Bigg]\Bigg>\Bigg|_{x=x^{\prime}}\;.
\end{eqnarray}

In order to evaluate this term we use the derivatives in
(\ref{67a}) and (\ref{69}) and we expand the remaining
exponentials of $\tilde{H}$ in $\tau$ up to terms in
$\lambda^{2}$. During the calculation the numerical coefficients
of the various terms can be evaluated by referring to the
following general integral
\begin{equation}\label{77}
I_{2}(\alpha,\beta)=\int\limits_{0}^{1}d\tau_{1}\tau_{1}^{\alpha}(1-\tau_{1})^{\beta}\;,
\end{equation}
where $(\alpha,\beta)$ are positive integers, and for which the
general solution in closed form is (see Appendix B)
\begin{equation}\label{78}
I_{2}(\alpha,\beta)=\frac{\Gamma(1+\alpha)\Gamma(1+\beta)}{\Gamma(2+\alpha+\beta)}\;.
\end{equation}

The explicit form of (\ref{76}) can be obtained with the help of
(\ref{78}) and the coincidence limits in Appendix A. After a
straightforward calculation one gets
\begin{eqnarray}\label{79}
\lefteqn{-\left<\textrm{tr}_{V}\;\Bigg[\int\limits_{0}^{1}d\tau_{1}e^{-(1-\tau_{1})\tilde{H}}
\bar{\mathcal{D}}_{\mu}a^{\mu\nu}\mathcal{D}_{\nu}
e^{-\tau_{1}\tilde{H}}\Bigg]\right>\Bigg|_{x=x^{\prime}}=}\nonumber\\
&&=\textrm{tr}_{V}\left(\Xi_{0}\right)+\lambda\textrm{tr}_{V}\left(\Xi_{1}\right)
+\lambda^{2}\textrm{tr}_{V}\left(\Xi_{2}\right)+O(\lambda^{3})\;,
\end{eqnarray}
where
\begin{eqnarray}\label{80}
\Xi_{0}&=&-\phi_{;\;\mu}{}^{;\;\mu}-\phi_{;\;\mu}\phi^{;\;\mu}\;,\\[3pt]
\label{800}
\Xi_{1}&=&0
% these terms are traceless
%-\sigma_{;\;\mu}{}^{;\;\mu}
%-h^{\mu\nu}{}_{;\;\mu}\phi_{;\;\nu}
%-h^{\mu\nu}\phi_{;\;\mu\nu}
%-2\phi^{;\;\nu}\sigma_{;\;\nu}
%-\phi_{;\;\mu}h^{\mu\nu}\phi_{;\;\nu}\nonumber\\
%&+&\frac{1}{2}h(\phi_{;\;\mu}{}^{;\;\mu}
%+\phi_{;\;\mu}\phi^{;\;\mu})
%-\frac{1}{4} h^{;\alpha}{}_{;\alpha}
\;,\\[3pt]
\label{801}
\Xi_{2}&=& \frac{1}{8}h h^{;\alpha}{}_{;\alpha}
+\frac{1}{4}h^{\mu\nu}h_{\mu\nu}^{;\alpha}{}_{;\alpha}
+\frac{1}{12}h_{;\;\mu}h^{;\;\mu}+\frac{1}{6}h_{\mu\nu;\;\rho}h^{\mu\nu;\;\rho}-\frac{1}{4}h^{\mu\nu}h_{;\;\mu\nu}\nonumber\\
&-&\frac{1}{4}h^{\mu\nu}{}_{;\;\mu}h_{;\;\nu}+\frac{1}{2}hh^{\mu\nu}{}_{;\;\mu}\phi_{;\;\nu}-h^{\mu\nu}{}_{;\;\mu}\sigma_{;\;\nu}-\frac{1}{8}h^{2}\phi_{;\;\mu}{}^{;\;\mu}-\frac{1}{4}h_{\mu\nu}h^{\mu\nu}\phi_{;\;\rho}{}^{;\;\rho}\nonumber\\
&+&\frac{1}{2}h\sigma_{;\;\mu}{}^{;\;\mu}+\frac{1}{2}hh^{\mu\nu}\phi_{;\;\mu\nu}-h^{\mu\nu}\sigma_{;\;\mu\nu}-\frac{1}{8}h^{2}\phi_{;\;\mu}\phi^{;\;\mu}-\frac{1}{4}h^{\mu\nu}h_{\mu\nu}\phi_{;\;\rho}\phi^{;\;\rho}\nonumber\\
&+&h\sigma_{;\;\mu}\phi^{;\;\mu}+\frac{1}{2}hh^{\mu\nu}\phi_{;\;\mu}\phi_{;\;\nu}-\sigma_{;\;\mu}\sigma^{;\;\mu}-2h^{\mu\nu}\sigma_{;\;\mu}\phi_{;\;\nu}\;.
\end{eqnarray}

In the notation of equation (\ref{53}) we can write, now, the
different contributions, in increasing order of $\lambda$, to the
coefficient $\tilde{a}_{1}$. In more details, by using the results
(\ref{64}), (\ref{73}), (\ref{80}) and recalling that
\begin{displaymath}
\tr_V\tilde{a}_{1}(\lambda)=\tr_V a_{1}^{(0)}
+\lambda \tr_V a_{1}^{(1)}+\lambda^{2}\tr_V a_{1}^{(2)}
+O(\lambda^{3})\;,
\end{displaymath}
we get
\begin{equation}\label{83}
a_{1}^{(0)}=\frac{1}{6}R-\phi_{;\;\mu}{}^{;\;\mu}-\phi_{;\;\mu}\phi^{;\;\mu}-q\;.
\end{equation}
Moreover, by using (\ref{730}) and (\ref{800}) we obtain
\begin{eqnarray}\label{84}
a_{1}^{(1)}&=&0
% these terms are traceless
%-\sigma_{;\;\mu}{}^{;\;\mu}
%-h^{\mu\nu}{}_{;\;\mu}\phi_{;\;\nu}
%-h^{\mu\nu}\phi_{;\;\mu\nu}
%-2\phi^{;\;\nu}\sigma_{;\;\nu}
%-\phi_{;\;\mu}h^{\mu\nu}\phi_{;\;\nu}
%-\frac{1}{6}h^{\mu\nu}{}_{;\;\mu\nu}\nonumber\\
%&+&\frac{1}{2}h(\phi_{;\;\mu}{}^{;\;\mu}
%+\phi_{;\;\mu}\phi^{;\;\mu})
%-\frac{1}{12} h^{;\alpha}{}_{;\alpha}
%+\frac{1}{6}h^{\mu\nu}R_{\mu\nu}
%-\frac{1}{12}hR
\;.
\end{eqnarray}
Finally, by combining the results in (\ref{64}), (\ref{731}) and
(\ref{801}), we have the following expression for the term of
order $\lambda^{2}$ in $\tilde{a}_{1}$, i.e.
\begin{eqnarray}\label{85}
a_{1}^{(2)}&=&-h^{\mu\nu}{}_{;\;\mu}\sigma_{;\;\nu}-h^{\mu\nu}\sigma_{;\;\mu\nu}-\sigma^{;\;\nu}\sigma_{;\;\nu}-\sigma_{;\;\mu}h^{\mu\nu}\phi_{;\;\nu}-\phi_{;\;\nu}h^{\mu\nu}\sigma_{;\;\mu}-\frac{1}{12}h^{\mu\nu}h_{;\;\mu\nu}\nonumber\\
&+&\frac{1}{2}\sigma_{;\;\mu}{}^{;\;\mu}h+\frac{1}{2}hh^{\mu\nu}{}_{;\;\mu}\phi_{;\;\nu}+h\phi_{;\;\mu}\sigma^{;\;\mu}+\frac{1}{2}hh^{\mu\nu}\phi_{;\;\mu\nu}+\frac{1}{2}hh^{\mu\nu}\phi_{;\;\mu}\phi_{;\;\nu}-\frac{1}{12}h^{\mu\nu}{}_{;\;\mu}h_{;\;\nu}\nonumber\\
&+&\frac{1}{12}h^{\mu\nu}{}_{;\;\mu\nu}h
+\frac{1}{12}h^{\mu\nu;\;\rho}h_{\mu\rho;\;\nu}
+\frac{1}{24}h h^{;\alpha}{}_{;\alpha}
+\frac{1}{12}h^{\mu\nu} h_{\mu\nu}{}^{;\alpha}{}_{;\alpha}
+\frac{1}{24}h^{\mu\nu;\;\rho}h_{\mu\nu;\;\rho}\nonumber\\
&-&\frac{1}{8}h^{2}q-\frac{1}{8}h^{2}\phi_{;\;\mu}{}^{;\;\mu}-\frac{1}{8}h^{2}\phi_{;\;\mu}\phi^{;\;\mu}-\frac{1}{4}h^{\mu\nu}h_{\mu\nu}q-\frac{1}{4}h^{\mu\nu}h_{\mu\nu}\phi_{;\;\rho}{}^{;\;\rho}\nonumber\\
&-&\frac{1}{4}h^{\mu\nu}h_{\mu\nu}\phi_{;\;\rho}\phi^{;\;\rho}+\frac{1}{48}h_{;\;\mu}h^{;\;\mu}-\frac{1}{12}hh^{\mu\nu}R_{\mu\nu}-\frac{1}{6}h^{\mu}{}_{\rho}h^{\nu\rho}R_{\mu\nu}+\frac{1}{48}h^{2}R\nonumber\\
&+&\frac{1}{24}h^{\mu\nu}h_{\mu\nu}R+\frac{1}{2}h\Theta\;.
\end{eqnarray}

%=================================================================
\section{Construction of the Action}
\setcounter{equation}0

In order to write the action of Spectral matrix Gravity, we need
to evaluate the global heat kernel coefficients $A_{0}$ and
$A_{1}$. As we already mentioned above, the coefficients $A_{k}$
are expressed in terms of integrals of the local heat kernel
coefficients $a_{k}$ (which are densities) or the coefficients
$\tilde{a}_{k}$ (which are scalars). Namely,
\begin{equation}\label{86}
A_{k}=\int\limits_{M}dx\;g^{\frac{1}{2}}\;\textrm{tr}_{V}(\tilde{a}_{k})\;.
\end{equation}
By using the equation (\ref{59}), we get
\begin{equation}\label{87}
A_{0}=\int\limits_{M}dx\;g^{\frac{1}{2}}\;\textrm{tr}_{V}\left[\mathbb{I}+\frac{\lambda^{2}}{8}(h^{2}+2h_{\mu\nu}h^{\mu\nu})\right]+O(\lambda^{3})\;.
\end{equation}

Now we use the equations (\ref{53}) and (\ref{83})-(\ref{85}) to
compute the coefficient $A_{1}$. By integrating by parts and by
noticing that the trace of a commutator of any two matrices
vanishes, up to terms of order $\lambda^{2}$, we obtain
\begin{eqnarray}\label{88}
A_{1}&=&\int\limits_{M}dx\;g^{\frac{1}{2}}\;\textrm{tr}_{V}\bigg\{-q-\phi^{;\;\mu}\phi_{;\;\mu}+\frac{1}{6}R+\lambda^{2}\bigg(-\sigma^{;\;\mu}\sigma_{;\;\mu}+\frac{1}{2}h\Theta\nonumber\\
&-&\frac{1}{2}h^{;\;\mu}\sigma_{;\;\mu}+\frac{1}{2}hh^{\mu\nu}\phi_{;\;\mu}\phi_{;\;\nu}+h\sigma^{;\;\nu}\phi_{;\;\nu}-\frac{1}{2}h_{;\;\mu}h^{\mu\nu}\phi_{;\;\nu}-2\sigma_{;\;\mu}h^{\mu\nu}\phi_{;\;\nu}\nonumber\\
&-&\frac{1}{12}hh^{\mu\nu}R_{\mu\nu}-\frac{1}{6}h^{\nu}{}_{\rho}h^{\rho\mu}R_{\mu\nu}-\frac{1}{12}h^{\mu\nu}{}_{;\;\mu}h_{;\;\nu}+\frac{1}{12}h^{\mu\nu;\;\rho}h_{\mu\rho;\;\nu}+\frac{1}{48}h^{2}R\nonumber\\
&+&\frac{1}{24}h^{\mu\nu}h_{\mu\nu}R-\frac{1}{48}h_{;\;\mu}h^{;\;\mu}-\frac{1}{24}h^{\mu\nu;\;\rho}h_{\mu\nu;\;\rho}-\frac{1}{8}h^{2}q-\frac{1}{8}h^{2}\phi_{;\;\mu}{}^{;\;\mu}\nonumber\\
&-&\frac{1}{8}h^{2}\phi_{;\;\mu}\phi^{;\;\mu}-\frac{1}{4}h^{\mu\nu}h_{\mu\nu}q-\frac{1}{4}h^{\mu\nu}h_{\mu\nu}\phi_{;\;\rho}{}^{;\;\rho}-\frac{1}{4}h^{\mu\nu}h_{\mu\nu}\phi_{;\;\rho}\phi^{;\;\rho}\bigg)\bigg\}\nonumber\\
&+&O(\lambda^{3})\;.
\end{eqnarray}

The invariant action functional is written as linear combination
of the coefficients $A_{0}$ and $A_{1}$ as shown in (\ref{2})
\begin{eqnarray}\label{89}
S&=&\frac{1}{16\pi
G}\int\limits_{M}dx\;g^{\frac{1}{2}}\bigg\{-6q-6\phi^{;\;\mu}\phi_{;\;\mu}+R-2\Lambda\nonumber\\
&+&\frac{\lambda^{2}}{N}\textrm{tr}_V
\bigg(-6\sigma^{;\;\mu}\sigma_{;\;\mu}-3h^{;\;\mu}\sigma_{;\;\mu}+3h\Theta+3hh^{\mu\nu}\phi_{;\;\mu}\phi_{;\;\nu}+6h\sigma^{;\;\nu}\phi_{;\;\nu}\nonumber\\
&-&3h^{\mu\nu}h_{;\;\mu}\phi_{;\;\nu}-12\sigma_{;\;\nu}h^{\mu\nu}\phi_{;\;\mu}-\frac{1}{2}hh^{\mu\nu}R_{\mu\nu}+\frac{1}{2}h^{\mu\nu}h^{\rho\sigma}R_{\sigma\mu\rho\nu}-\frac{1}{2}h^{\nu}{}_{\rho}h^{\rho\mu}R_{\mu\nu}\nonumber\\[3pt]
&-&\frac{1}{2}h^{\mu\nu}{}_{;\;\mu}h_{;\;\nu}+\frac{1}{2}h^{\mu\nu}{}_{;\;\nu}h_{\mu\rho}{}^{;\;\rho}+\frac{1}{8}h^{2}R+\frac{1}{4}h^{\mu\nu}h_{\mu\nu}R-\frac{1}{8}h_{;\;\mu}h^{;\;\mu}-\frac{1}{4}h^{\mu\nu;\;\rho}h_{\mu\nu;\;\rho}\nonumber\\[3pt]
&-&\frac{3}{4}h^{2}q-\frac{3}{4}h^{2}\phi_{;\;\mu}{}^{;\;\mu}-\frac{3}{4}h^{2}\phi_{;\;\mu}\phi^{;\;\mu}-\frac{3}{2}h^{\mu\nu}h_{\mu\nu}q-\frac{3}{2}h^{\mu\nu}h_{\mu\nu}\phi_{;\;\rho}{}^{;\;\rho}\nonumber\\[3pt]
&-&\frac{3}{2}h^{\mu\nu}h_{\mu\nu}\phi_{;\;\rho}\phi^{;\;\rho}-\frac{\Lambda}{4}h^{2}-\frac{\Lambda}{2}h_{\mu\nu}h^{\mu\nu}\bigg)\bigg\}\nonumber\\
&+&O(\lambda^{3})\;.
\end{eqnarray}
It is straightforward to show that the action functional that we
obtained is invariant under the diffeomorphisms
and the gauge transformation $h^{\mu\nu}\rightarrow
Uh^{\mu\nu}U^{-1}$.

The next task is to find the equations of motion for the fields
$\sigma$, $h_{\mu\nu}$ and $\phi$ by varying the action
functional. In this way we will explicitly find the noncommutative
corrections to Einstein's equations.

%=================================================================
\section{The Equations of Motion}
\setcounter{equation}0

By performing the variation with respect to the field $\sigma$, we
obtain the equation
\begin{equation}\label{90}
4\Box\sigma+\Box
h-2h\Box\phi-2h^{;\;\nu}\phi_{;\;\nu}+4h^{\mu\nu}{}_{;\;\nu}\phi_{;\;\mu}+4h^{\mu\nu}\phi_{;\;\mu\nu}+O(\lambda^{3})=0\;.
\end{equation}
Here $\Box$ is the Laplacian in the Euclidean case and the D'Alambertian in the
pseudo-Euclidean case.
For the matrix-valued field $h^{\mu\nu}$ we obtain the
equation
\begin{eqnarray}\label{92}
\lefteqn{g^{\mu\nu}\Box\sigma+g^{\mu\nu}\Theta+h\phi^{;\;(\mu}\phi^{;\;\nu)}+g^{\mu\nu}h^{\rho\sigma}\phi_{;\;\rho}\phi_{;\;\sigma}+2g^{\mu\nu}\sigma^{;\;\rho}\phi_{;\;\rho}}\nonumber\\
&-&h^{;(\mu}\phi^{;\;\nu)}-4\sigma^{;(\mu}\phi^{;\;\nu)}+g^{\mu\nu}h^{\rho\sigma}{}_{;\;\rho}\phi_{;\;\sigma}+g^{\mu\nu}h^{\rho\sigma}\phi_{;\;\rho\sigma}-\frac{1}{6}g^{\mu\nu}h^{\rho\sigma}R_{\rho\sigma}\nonumber\\
&-&\frac{1}{6}hR^{\mu\nu}+\frac{1}{3}h_{\rho\sigma}R^{\sigma\mu\rho\nu}+\frac{1}{3}h^{\rho(\mu}R^{\nu)}{}_{\rho}+\frac{1}{6}h^{;\;(\mu\nu)}+\frac{1}{6}g^{\mu\nu}h^{\rho\sigma}{}_{;\;\rho\sigma}-\frac{1}{3}h^{(\mu}{}_{\rho}{}^{;\;|\rho|\nu)}\nonumber\\
&+&\frac{1}{12}g^{\mu\nu}hR+\frac{1}{6}h^{\mu\nu}R+\frac{1}{12}g^{\mu\nu}\Box
h+\frac{1}{6}\Box h^{\mu\nu}-\frac{1}{2}g^{\mu\nu}hq-\frac{1}{2}g^{\mu\nu}h\Box\phi\nonumber\\
&-&\frac{1}{2}g^{\mu\nu}h\phi_{;\;\rho}\phi^{;\;\rho}-h^{\mu\nu}q-h^{\mu\nu}\Box\phi-h^{\mu\nu}\phi_{;\;\rho}\phi^{;\;\rho}-\frac{\Lambda}{6}g^{\mu\nu}h-\frac{\Lambda}{3}h^{\mu\nu}+O(\lambda^{3})=0\;.\nonumber\\
\phantom{2}
\end{eqnarray}

The variation of the action with respect to the scalar field
$\phi$ yields
\begin{eqnarray}\label{91}
4\Box\phi&=&-\frac{\lambda^{2}}{N}\textrm{tr}_{V}\bigg(-2hh^{\mu\nu}\phi_{;\;\mu\nu}-2h_{;\;\mu}h^{\mu\nu}\phi_{;\;\nu}-2hh^{\mu\nu}{}_{;\;\mu}\phi_{;\;\nu}\nonumber\\
&-&2h_{;\;\nu}\sigma^{;\;\nu}-2h\Box\sigma+h^{\mu\nu}{}_{;\;\nu}h_{;\;\mu}+h^{\mu\nu}h_{;\;\mu\nu}+4\sigma_{;\;\mu\nu}h^{\mu\nu}+4\sigma_{;\;\nu}h^{\mu\nu}{}_{;\;\mu}\nonumber\\
&-&\frac{1}{2}h\Box h-\frac{1}{2}h_{;\;\mu}h^{;\;\mu}+hh_{;\;\mu}\phi^{;\;\mu}+\frac{1}{2}h^{2}\Box\phi-h^{\mu\nu}\Box h_{\mu\nu}-h_{\mu\nu;\;\rho}h^{\mu\nu;\;\rho}\nonumber\\
&+&2h_{\mu\nu}h^{\mu\nu;\;\rho}\phi_{;\;\rho}+h^{\mu\nu}h_{\mu\nu}\Box\phi\bigg)+O(\lambda^{3})\;.
\end{eqnarray}

The equation of motion for the field $g^{\mu\nu}$ can be written
in the following form
\begin{equation}\label{93}
R^{\mu\nu}-\frac{1}{2}g^{\mu\nu}R+\Lambda
g^{\mu\nu}=T^{\mu\nu}+\frac{\lambda^{2}}{N}\textrm{tr}_{V}\mathscr{A}^{\mu\nu}\;.
\end{equation}
Here the tensor $T^{\mu\nu}$ is
\begin{equation}\label{94}
T^{\mu\nu}=6\phi^{;\;(\mu}\phi^{;\;\nu)}-3g^{\mu\nu}\phi_{;\;\rho}\phi^{;\;\rho}-3qg^{\mu\nu}\;,
\end{equation}
which represents the stress-energy tensor for a massless scalar
field. The tensor $\mathcal{A}^{\mu\nu}$ represents, instead, the
stress-energy tensor for the fields $h^{\mu\nu}$ and $\sigma$.

The equation (\ref{93}) is the main result of our paper. As we can
see, the new fields of our model, $h^{\mu\nu}$ and $\sigma$,
contribute to modify the standard Einstein equations. More
precisely they contribute to an additional term in the
stress-energy tensor.

The tensor $\mathscr{A}^{\mu\nu}$ can be written as the sum of six
terms:
\begin{equation}
\mathscr{A}^{\mu\nu}=\mathscr{A}^{\mu\nu}_{(1)}+\mathscr{A}^{\mu\nu}_{(2)}+\mathscr{A}^{\mu\nu}_{(3)}+\mathscr{A}^{\mu\nu}_{(4)}+\mathscr{A}^{\mu\nu}_{(5)}+\mathscr{A}^{\mu\nu}_{(6)}\;.
\end{equation}
In the first term we have only derivatives of the field $\sigma$
\begin{eqnarray}\label{95}
\mathscr{A}^{\mu\nu}_{(1)}&=&6\sigma^{;\;(\mu}\sigma^{;\;\nu)}-3g^{\mu\nu}\sigma_{;\;\rho}\sigma^{;\;\rho}-6h\sigma^{;\;(\mu}\phi^{;\;\nu)}+3\sigma^{;\;(\mu}h^{;\;\nu)}+6h^{\mu\nu}\sigma^{;\;\rho}\phi_{;\;\rho}\nonumber\\
&+&3\sigma_{;\;\rho}{}^{;\;\rho}h^{\mu\nu}+3g^{\mu\nu}\left(h\sigma^{;\;\rho}\phi_{;\;\rho}-\frac{1}{2}\sigma^{;\;\rho}h_{;\;\rho}-2\sigma_{;\;\rho}\phi_{;\;\sigma}h^{\rho\sigma}\right)\;.
\end{eqnarray}
The second term only contains derivatives of the scalar field
$\phi$, namely
\begin{eqnarray}\label{96}
\mathscr{A}^{\mu\nu}_{(2)}&=&3h^{\rho\tau}\phi_{;\;\rho}\phi_{;\;\tau}\left(h^{\mu\nu}+\frac{1}{2}g^{\mu\nu}h\right)+3\phi_{;\;\tau}\left(h^{\rho\tau}{}_{;\;\rho}h^{\mu\nu}-\frac{1}{2}g^{\mu\nu}h^{\rho\tau}h_{;\;\rho}\right)\nonumber\\
&-&\frac{3}{2}(\phi_{;\;\rho}\phi^{;\;\rho}+\phi_{;\;\rho}{}^{;\;\rho})\left(2h^{(\mu}{}_{\sigma}h^{\nu)\sigma}+hh^{\mu\nu}+\frac{1}{4}h^{2}g^{\mu\nu}+\frac{1}{2}g^{\mu\nu}h^{\alpha\beta}h_{\alpha\beta}\right)\nonumber\\
&+&3h^{\rho\sigma}h^{\mu\nu}\phi_{;\;\rho\sigma}+\frac{3}{2}(\phi^{;\;(\mu\nu)}+\phi^{;\;(\mu}\phi^{;\;\nu)})\left(h_{\rho\sigma}h^{\rho\sigma}+\frac{1}{2}h^{2}\right)\;.
\end{eqnarray}
The third term only contains second derivatives of the
matrix-valued tensor field $h^{\mu\nu}$,
\begin{eqnarray}\label{96a}
\mathscr{A}^{\mu\nu}_{(3)}&=&h^{\alpha(\mu}{}_{;\;(\alpha\sigma)}h^{\nu)\sigma}+\frac{1}{2}h^{\alpha\sigma}{}_{;\;\alpha\sigma}h^{\mu\nu}-\frac{1}{2}hh^{\alpha(\mu;\;\nu)}{}_{;\;\alpha}+\frac{1}{4}g^{\mu\nu}hh^{\rho\sigma}{}_{;\;\rho\sigma}\nonumber\\
&+&\frac{1}{2}h^{\sigma(\mu}h^{\nu)\rho}{}_{;\;\rho\sigma}-\frac{1}{2}h^{\rho\sigma}h^{\mu\nu}{}_{;\;\rho\sigma}-h_{\rho\sigma}{}^{;\;(\nu|\rho|}h^{\mu)\sigma}+h^{(\nu}{}_{\rho}h^{\mu)\rho;\;\sigma}{}_{;\;\sigma}\nonumber\\
&+&\frac{1}{4}hh^{\mu\nu;\;\sigma}{}_{;\;\sigma}-\frac{1}{4}g^{\mu\nu}h^{\sigma}{}_{\rho}h^{\rho\alpha}{}_{;\;[\alpha\sigma]}+\frac{1}{2}h^{\rho\sigma}h_{\rho\sigma}{}^{;\;(\mu\nu)}-\frac{1}{2}g^{\mu\nu}h^{\rho\sigma}h_{\rho\sigma;\;\alpha}{}^{;\;\alpha}\;.\nonumber\\
\phantom{2}
\end{eqnarray}
The fourth term contains only find first derivatives of
$h^{\mu\nu}$, namely
\begin{eqnarray}\label{96b}
\mathscr{A}^{\mu\nu}_{(4)}&=&-\frac{1}{2}h^{\rho\sigma}{}_{;\;\rho}h^{\mu\nu}{}_{;\;\sigma}+\frac{1}{2}h^{;\;\rho}h^{\mu\nu}{}_{;\;\rho}-\frac{1}{2}h^{\rho(\mu}{}_{;\;(\rho}h^{\nu)\lambda}{}_{;\;\lambda)}+\frac{1}{2}g^{\mu\nu}h_{;\;\sigma}h^{\rho\sigma}{}_{;\;\rho}\nonumber\\
&+&\frac{1}{2}h^{\sigma(\mu}{}_{;\;\rho}h^{\nu)\rho}{}_{;\;\sigma}+\frac{1}{4}h^{\rho\sigma;\;(\mu}h_{\rho\sigma}{}^{;\;\nu)}-h_{\sigma\rho}{}^{;\;(\nu}h^{\mu)\sigma;\;\rho}+h^{\rho(\nu;\;|\tau|}h_{\rho}{}^{\mu)}{}_{;\;\tau}\nonumber\\
&+&\frac{1}{2}h^{;\;(\nu}h^{\mu)\rho}{}_{;\;\rho}+\frac{1}{4}g^{\mu\nu}h_{\rho\sigma;\;\alpha}h^{\rho\alpha;\;\sigma}-\frac{1}{2}h_{;\;\rho}h^{\rho(\mu;\;\nu)}-\frac{5}{8}g^{\mu\nu}h^{\rho\sigma;\;\alpha}h_{\rho\sigma;\;\alpha}\;.\nonumber\\
\phantom{}
\end{eqnarray}
The fifth coefficient contains only first and second derivatives
of $h$
\begin{eqnarray}\label{97}
\mathscr{A}^{\mu\nu}_{(5)}&=&\frac{1}{4}h^{;\;\tau}{}_{;\;\tau}h^{\mu\nu}+\frac{1}{4}h^{;\;(\mu\nu)}h+\frac{3}{8}h^{;\;(\mu}h^{;\;\nu)}\nonumber\\
&-&\frac{1}{2}g^{\mu\nu}\left(\frac{1}{2}hh_{;\;\rho}{}^{;\;\rho}-h_{;\;\rho\sigma}h^{\rho\sigma}+\frac{5}{8}h^{;\;\rho}h_{;\;\rho}\right)\;.
\end{eqnarray}
The last term, $\mathcal{A}^{\mu\nu}_{(6)}$, does not contain any
derivative of $h^{\mu\nu}$, namely
\begin{eqnarray}\label{98}
\mathscr{A}^{\mu\nu}_{(6)}&=&3\Theta\left(h^{\mu\nu}+\frac{1}{2}g^{\mu\nu}h\right)-\frac{\Lambda}{2}\left[\left(h^{\mu\nu}+\frac{1}{4}g^{\mu\nu}h\right)h+2h^{(\nu}{}_{\rho}h^{\mu)\rho}+\frac{1}{2}g^{\mu\nu}h^{\rho\sigma}h_{\rho\sigma}\right]\nonumber\\
&-&\frac{1}{2}R_{\alpha\beta}\Bigg[h^{\alpha(\nu}h^{\mu)\beta}+h^{\mu\nu}h^{\alpha\beta}+g^{\mu\nu}\left(hh^{\alpha\beta}+h^{\alpha}{}_{\rho}h^{\beta\rho}\right)+h^{2}g^{\alpha(\mu}g^{\nu)\beta}\nonumber\\
&+&h^{\rho\sigma}h_{\rho\sigma}g^{\alpha(\mu}g^{\nu)\beta}\Bigg]+\frac{1}{2}h^{\alpha\sigma}h^{\rho(\mu}R^{\nu)}{}_{\alpha\rho\sigma}+\frac{1}{4}g^{\mu\nu}h^{\rho\sigma}h^{\alpha\beta}R_{\beta\rho\alpha\sigma}\nonumber\\
&+&\frac{1}{8}(R-6q)\left(g^{\mu\nu}h_{\rho\sigma}h^{\rho\sigma}+4h_{\rho}{}^{(\mu}h^{\nu)\rho}\right)+\frac{1}{16}(R-6q)\left(g^{\mu\nu}h^{2}+4hh^{\mu\nu}\right)\;.\nonumber\\
\phantom{}
\end{eqnarray}

The dynamics described by the equations (\ref{90}), (\ref{92}),
(\ref{91}) and (\ref{93}) can be studied by using an iterative
method. Let us write the solution for the background fields $\phi$
and $g^{\mu\nu}$ as Taylor expansion in the deformation parameter
$\lambda$ as follows
\begin{eqnarray}\label{willy}
\phi&=&\phi_{0}+\lambda\phi_{1}+\lambda^{2}\phi_{2}+O(\lambda^{3})\;,\nonumber\\
g^{\mu\nu}&=&g^{\mu\nu}_{0}+\lambda
g^{\mu\nu}_{1}+\lambda^{2}g^{\mu\nu}_{2}+O(\lambda^{3})\;.
\end{eqnarray}
By substituting these expressions in equations (\ref{91}) and
(\ref{93}) we obtain, for the terms of order $\lambda^{0}$, the
dynamical equations
\begin{eqnarray}\label{willy1}
\Box\phi&=&0\;,\nonumber\\
R^{\mu\nu}-\frac{1}{2}g^{\mu\nu}R+\Lambda g^{\mu\nu}&=&0\;.
\end{eqnarray}
As we can see from the last equations the term $g^{\mu\nu}_{0}$ is
nothing but the solution of the ordinary Einstein equation in
vacuum with cosmological constant. By substituting the solutions
to (\ref{willy1}) back into the equations of motion for the fields
$\sigma$ and $h^{\mu\nu}$ we get equations of the form
\begin{eqnarray}\label{willy2}
\Phi_{1}(g^{\mu\nu}_{0},\phi_{0})\sigma&=&O(\lambda^{2})\;,\nonumber\\
\Phi_{2}(g^{\mu\nu}_{0},\phi_{0})h^{\mu\nu}&=&O(\lambda^{2})\;,
\end{eqnarray}
where $\Phi_{1}(g^{\mu\nu}_{0},\phi_{0})$ and
$\Phi_{2}(g^{\mu\nu}_{0},\phi_{0})$ are linear second order
partial differential operators.

By iterating this process we can, in principle, find the solution
to our dynamical equations in form of a Taylor series in
$\lambda$.

%=================================================================
\section{Spectrum of Matrix Gravity on De Sitter Space}
\setcounter{equation}0

The action for matrix gravity obtained in the previous section is
a functional of the fields $\phi$, $\sigma$, $h^{\mu\nu}$ and
$g^{\mu\nu}$. The dynamics is described by a system of non-linear
partial differential equations coupled with each other. We
analyze, now, the dynamics of the theory. For simplicity we will
set, from now on, $Q=0$. This particular value for the
matrix-valued scalar $Q$ will not affect our analysis.

As already mentioned above, from the equation of motion (\ref{91})
for the field $\phi$, we can see that to the zeroth order in the
deformation parameter $\lambda$ the field $\phi$ satisfies the
following equation
\begin{equation}\label{99}
\Box\phi=0\;.
\end{equation}
As it is well known, the solution of the last equation represents
a wave propagating in the whole space. Since we require that
$\phi$ vanishes at infinity, the only solution is
$\phi=O(\lambda^{2})$ in the whole space.

With this solution for the field $\phi$, the matrix-valued
function $\rho$ defined in (\ref{39}) becomes
\begin{equation}\label{100}
\rho=e^{\lambda\sigma}\;.
\end{equation}
A deeper analysis shows that the matrix-valued scalar field
$\sigma$ is not an independent field. Following
\cite{avramidi04,avramidi04b} the general form of $\rho$ can be
written as
\begin{equation}\label{101}
\rho=\omega^{-\frac{1}{4}}\;,
\end{equation}
where
\begin{equation}\label{102}
\omega=-\frac{1}{m!}\varepsilon_{\mu_{1}\ldots\mu_{m}}\varepsilon_{\nu_{1}\ldots\nu_{m}}a^{\mu_{1}\nu_{1}}\cdots
a^{\mu_{m}\nu_{m}}\;.
\end{equation}
By using the decomposition (\ref{38}) of $h^{\mu\nu}$ in equation
(\ref{102}) we get the following formula, up to the term linear in
the deformation parameter $\lambda$,
\begin{equation}\label{103}
\sigma=-\frac{1}{4}h+\frac{\lambda}{8}h_{\mu\nu}h^{\mu\nu}+O(\lambda^{2})\;.
\end{equation}
We can write down, now, the action by imposing the constraints
$\phi=O(\lambda^{2})$ and (\ref{103}). The final result is the
following
\begin{eqnarray}\label{104}
S&=&\frac{1}{16\pi
G}\int\limits_{M}dx\;g^{\frac{1}{2}}\bigg\{-6\phi^{;\;\mu}\phi_{;\;\mu}+R-2\Lambda+\frac{\lambda^{2}}{N}\textrm{tr}_{V}\bigg[\frac{1}{4}g^{\mu\nu}h_{;\;\mu}h_{;\;\nu}\nonumber\\
&-&\frac{1}{2}hh^{\mu\nu}R_{\mu\nu}+\frac{1}{2}h^{\mu\nu}h^{\rho\sigma}R_{\sigma\mu\rho\nu}-\frac{1}{2}h^{\nu}{}_{\rho}h^{\rho\mu}R_{\mu\nu}-\frac{1}{2}h^{\mu\nu}{}_{;\;\mu}h_{;\;\nu}\nonumber\\
&+&\frac{1}{2}h^{\mu\nu}{}_{;\;\nu}h_{\mu\rho}{}^{;\;\rho}+\frac{1}{8}h^{2}R+\frac{1}{4}h^{\mu\nu}h_{\mu\nu}R-\frac{1}{4}h^{\mu\nu;\;\rho}h_{\mu\nu;\;\rho}\nonumber\\
&-&\frac{\Lambda}{4}h^{2}-\frac{\Lambda}{2}h_{\mu\nu}h^{\mu\nu}\bigg)\bigg\}+O(\lambda^{3})\;.
\end{eqnarray}

The action depends, now, only on the independent tensor fields
$g^{\mu\nu}$, $h^{\mu\nu}$ and the scalar field $\phi$. Therefore,
we will have only two equations that describe the dynamics of the
theory. These dynamical equations can be easily derived from the
ones given in the previous section by imposing the conditions
(\ref{103}) and $\phi=O(\lambda^{2})$.

The action (\ref{104}) and the equations of motions for the fields
evaluated in the previous section, assume a simple form on
maximally symmetric background geometries. As we mentioned in the
previous section, the $\lambda^{0}$ term of the background field
$g^{\mu\nu}_{0}$ is solution of the Einstein equations in vacuum
with cosmological constant (\ref{willy1}). In this section we
consider the De Sitter solution to the equation (\ref{willy1}). In
this maximally symmetric case the Ricci and Riemann tensors take
the following form
\begin{equation}\label{110}
R^{\mu}{}_{\nu\alpha\beta}=\frac{1}{n(n-1)}(\delta^{\mu}{}_{\alpha}g_{\nu\beta}-\delta^{\mu}{}_{\beta}g_{\nu\alpha})R\qquad\textrm{and}\qquad
R_{\mu\nu}=\frac{1}{n}g_{\mu\nu}R\;.
\end{equation}
The De Sitter metric gives a solution of the classical equations
provided
\begin{equation}\label{110a}
R=\frac{2n}{n-2}\Lambda\;.
\end{equation}

By substituting the expressions in equation (\ref{110}) in the
action (\ref{104}), we find a form of the action functional valid
in De Sitter geometry, namely
\begin{eqnarray}\label{111}
&\phantom{2}&S=\frac{1}{16\pi
G}\int\limits_{M}dx\;g^{\frac{1}{2}}\bigg\{-6\phi^{;\;\mu}\phi_{;\;\mu}+R-2\Lambda\nonumber\\
&+&\frac{\lambda^{2}}{N}\textrm{tr}_{V}\bigg[\frac{1}{4}h(-\Box+\mu_{1})h
-\frac{1}{4}h_{\mu\nu}(-\Box+\mu_{2})h^{\mu\nu}+\frac{1}{2}h^{\mu\nu}{}_{;\;\nu}h_{\rho\mu}{}^{;\;\rho}-\frac{1}{2}h^{\mu\nu}{}_{;\;\mu}h_{;\;\nu}\bigg]\bigg\}\nonumber\\
&+&O(\lambda^{3})\;,
\end{eqnarray}
where the terms $\mu_{1}$ and $\mu_{2}$ are defined as follows
\begin{eqnarray}\label{112}
\mu_{1}&=&\frac{n^{2}-5n+8}{2n(n-1)}R-\Lambda\;,\\
\mu_{2}&=&-\frac{n-3}{n-1}R+2\Lambda\;.
\end{eqnarray}

It is interesting, at this point of the discussion, to derive
explicitly the spectrum of the theory. In order to achieve this
result we need to decompose the field $h_{\mu\nu}$ in its
irreducible modes: traceless transverse tensor mode, transverse
vector mode, scalar mode and trace part. In other words we can
write $h_{\mu\nu}$ as
\begin{equation}\label{112b}
h_{\mu\nu}=\bar{h}_{\mu\nu}^{\bot}+\frac{1}{n}g_{\mu\nu}\varphi+2\zeta_{(\mu;\;\nu)}^{\bot}+\psi_{;\;\mu\nu}\;,
\end{equation}
where the scalar field $\varphi$ is defined as follows
\begin{displaymath}
\varphi=h-\Box\psi\;,
\end{displaymath}
and the fields $\bar{h}_{\mu\nu}^{\bot}$ and $\zeta_{\mu}^{\bot}$
satisfy the conditions
\begin{equation}\label{112a}
\nabla^{\mu}\bar{h}_{\mu\nu}^{\bot}=0\;,\qquad
g^{\mu\nu}\bar{h}_{\mu\nu}^{\bot}=0\;,\qquad\nabla^{\mu}\zeta_{\mu}^{\bot}=0\;.
\end{equation}
We can now substitute the expression (\ref{112}) in the action
(\ref{111}), and evaluate the terms separately. Explicitly we
obtain
\begin{eqnarray}\label{113}
\int\limits_{M}dx\;g^{\frac{1}{2}}\left(\frac{1}{2}h^{\mu\nu}{}_{;\;\nu}h_{\rho\mu}{}^{;\;\rho}\right)&=&\int\limits_{M}dx\;g^{\frac{1}{2}}\Bigg[-\frac{1}{2n^{2}}\varphi\Box\varphi-\frac{1}{n}\varphi\left(\Box+\frac{R}{n}\right)\Box\psi\nonumber\\
&+&\frac{1}{2}\zeta^{\bot}_{\mu}\left(\Box+\frac{R}{n}\right)^{2}\zeta^{\bot\mu}-\frac{1}{2}\psi\left(\Box+\frac{R}{n}\right)^{2}\Box\psi\Bigg]\;,
\end{eqnarray}
\begin{eqnarray}\label{114}
-\int\limits_{M}dx\;g^{\frac{1}{2}}\left(\frac{1}{2}h^{\mu\nu}{}_{;\;\mu}h_{;\;\nu}\right)&=&\int\limits_{M}dx\;g^{\frac{1}{2}}\Bigg[\frac{1}{2n}\varphi\Box\varphi+\frac{1}{2}\left(\frac{n+1}{n}\right)\varphi\left(\Box+\frac{R}{n+1}\right)\Box\psi\nonumber\\
&+&\frac{1}{2}\psi\left(\Box+\frac{R}{n}\right)\Box^{2}\psi\Bigg]\;,
\end{eqnarray}
\begin{eqnarray}\label{115}
\lefteqn{\int\limits_{M}dx\;g^{\frac{1}{2}}\;h_{\mu\nu}(-\Box+\mu_{2})h^{\mu\nu}=}\nonumber\\
&=&\int\limits_{M}dx\;g^{\frac{1}{2}}\Bigg\{\bar{h}_{\mu\nu}^{\bot}(-\Box+\mu_{2})\bar{h}^{\bot\mu\nu}+\frac{1}{n}\varphi(-\Box+\mu_{2})\varphi\nonumber\\
&+&\frac{2}{n}\varphi(-\Box+\mu_{2})\Box\psi+2\zeta^{\bot}_{\mu}\left(\Box+\frac{R}{n}\right)\left(\Box-\mu_{2}+\frac{n+1}{n(n-1)}R\right)\zeta^{\bot\mu}\nonumber\\
&-&\psi\left[\Box^{2}+\left(\frac{3R}{n}-\mu_{2}\right)\Box+\frac{R}{n}\left(\frac{2R}{n-1}-\mu_{2}\right)\right]\Box\psi\Bigg\}\;,
\end{eqnarray}
and finally
\begin{eqnarray}\label{116}
&\phantom{2}&\int\limits_{M}dx\;g^{\frac{1}{2}}\;h(-\Box+\mu_{1})h=\nonumber\\
&=&\int\limits_{M}dx\;g^{\frac{1}{2}}\Big[\varphi(-\Box+\mu_{1})\varphi+2\varphi(-\Box+\mu_{1})\Box\psi+\psi(-\Box+\mu_{1})\Box^{2}\psi\Big]\;.
\end{eqnarray}

By using the decompositions (\ref{113}) through (\ref{116}) we
rewrite the action (\ref{111}) in terms of the irreducible modes
of $h^{\mu\nu}$, namely
\begin{eqnarray}\label{117}
S&=&\frac{1}{16\pi
G}\int\limits_{M}dx\;g^{\frac{1}{2}}\Bigg\{-6\phi^{;\;\mu}\phi_{;\;\mu}
+R-2\Lambda+\frac{\lambda^{2}}{N}\textrm{tr}_{V}\Bigg[
-\frac{1}{4}\bar{h}_{\mu\nu}^{\bot}(-\Box+\mu_{2})\bar{h}^{\bot\mu\nu}
\nonumber\\
&+&\frac{(n-1)(n-2)}{4n^{2}}\;\varphi\left(-\Box+\frac{n}{2(n-1)}R
-\frac{n(n+2)}{(n-1)(n-2)}\Lambda\right)\varphi\nonumber\\
&-&\frac{1}{2}\zeta^{\bot}_{\mu}\left(\Box+\frac{R}{n}\right)
\left(\frac{n-2}{n}R-2\Lambda\right)\zeta^{\bot\mu}+\frac{n+2}{4n}
\varphi\left(\frac{n-2}{n}R-2\Lambda\right)\Box\psi\nonumber\\
&+&\frac{3}{8}\psi\left(\Box+\frac{2R}{3n}\right)\left(\frac{n-2}{n}R
-2\Lambda\right)\Box\psi\Bigg]\Bigg\}+O(\lambda^{3})\;.
\end{eqnarray}

It is straightforward to show now that on the mass shell,
(\ref{110a}), the terms containing the fields $\zeta^{\bot}_{\mu}$
and $\psi$ vanish identically.
More precisely we obtain the following form for the
action functional
\begin{eqnarray}\label{118}
S&=&\frac{1}{16\pi
G}\int\limits_{M}dx\;g^{\frac{1}{2}}\Bigg\{
\frac{4\Lambda}{n-2}
%\nonumber\\
%&+&
+\frac{\lambda^{2}}{N}\textrm{tr}_{V}\Bigg[-\frac{1}{4}
\bar{h}_{\mu\nu}^{\bot}\left(-\Box+\frac{4\Lambda}{(n-1)(n-2)}\right)
\bar{h}^{\bot\mu\nu}\nonumber\\
&+&\frac{(n-1)(n-2)}{4n^{2}}\;\varphi
\left(-\Box-\frac{2n\Lambda}{(n-1)(n-2)}\right)\varphi\Bigg]\Bigg\}
+O(\lambda^{3})\;.
\end{eqnarray}
Thus on the mass shell the only remaining fields are the (traceless)
matrix-valued traceless transverse tensor  $\bar{h}_{\mu\nu}^{\bot}$ and
the (traceless) matrix-valued scalar field $\varphi$. This action looks
exactly the same as in general relativity, the only difference being
that the fields are matrix-valued and traceless. Therefore, it describes
$(N-1)$ spin-2 particles and $(N-1)$ spin-0 particles. Note also, that
exactly as in general relativity the scalar conformal mode is
unstable.

%=================================================================
\section{Concluding Remarks}
\setcounter{equation}{0}

In this paper we studied a noncommutative deformation of general
relativity (called Spectral Matrix Gravity) proposed in
\cite{avramidi04b} where the noncommutative limit has been
explicitly evaluated. The approach of the paper \cite{avramidi04b}
to construct the action for Matrix Gravity differs from the one proposed
in
\cite{avramidi04,avramidi03,fucci08}. In the latter the action of
our model was a straightforward generalization of the
Einstein-Hilbert action in which the measure and the scalar
curvature were matrix-valued quantities. This last approach seems
to have some intrinsic arbitrariness due to the freedom of
choosing the particular form of the matrix-valued measure (for a
discussion see \cite{fucci08}). In order to avoid these issues, in
\cite{avramidi04b}  the action of matrix gravity was defined as a
linear combination of the first two global heat kernel
coefficients (\ref{2}) of a non-Laplace type partial differential
operator.

By using the covariant Fourier transform method we were able to
evaluate the coefficients $A_{0}$ and $A_{1}$, and as a result,
the action functional within the perturbation theory  in the
deformation parameter $\lambda$. The main result of this paper is
the derivation of the modified Einstein equations in (\ref{93}) in
the weak deformation limit. In this case the pure noncommutative
fields, namely $h^{\mu\nu}$ and $\sigma$, contribute to the
right-hand side of the Einstein equation, that is, the
stress-energy tensor. The explicit form of these non-commutative
correction terms has been derived in (\ref{95}) through (\ref{98})
for the first time.

It is worth noting that the dynamical equations for the fields,
found in this paper, are \emph{classical} and therefore they
should be studied from the classical point of view.  It is an
intriguing question to study the physical and mathematical effects
of the non-commutative corrections. First of all, this is the
problem of singularities in general relativity. Another very
interesting task would be to find some basic particular solutions,
in particular, static spherically symmetric solutions
(non-commutative Schwarzschild)  and time-dependent homogeneous
solutions (non-commutative Robertson-Walker). Finally, it is the
question whether the non-commutative fields could account for the
dark matter and the dark energy. All these questions require
further study.

Some of the physical implications of Matrix Gravity have been
extensively discussed in \cite{avramidi03,avramidi04,avramidi04b}.
This theory exhibits non-geodesic motion which can be related to
a violation of the equivalence principle, moreover, because of the new
gauge symmetry, there are new physical conserved charges. At last, this
theory represents a consistent model of interacting spin-2 particles
on curved space which usually was a problem. An interesting question is
the limit as $N\rightarrow\infty$ of our model, this might be related to
matrix models and string theory.

As it is outlined in the introduction Matrix Gravity can be considered
as a Gravitational Chromodynamics describing the gravitational interaction
of a new degree of freedom that we call {\it gravitational color}.
Whether or not
it is related to the color of QCD is an open question. Let's suppose
for simplicity that it is the same, and that the gauge group of Matrix Gravity
is nothing but $SU(3)$. If one pushes this
analogy with QCD
to its logical limit then this would mean that the theory predicts
that the {\it gravitational interaction of quarks depends on their colors}.
Exactly as in QCD the strong interaction between quark of color $i$ and
a quark of color $j$ is transmitted by gluon of type $(ij)$, the
gravitational interaction between quark of color $i$ and
a quark of color $j$ is transmitted by the graviton of type $(ij)$.
In this case, all particles in the electro-weak sector, including photon,
do not feel the gravitational color. In that  sense it is `{\it dark}'.
Notice that in the non-relativistic limit the Newtonian potential will
also become `matrix-valued'.
One can go even further. Since the usual (white) mass is determined by
the sum of the color masses, one can assume that the color masses can be even
negative. Then the gravitational interaction of such particles would include
non only attractive forces but also repellent forces (antigravity?).
This feature could then solve the mistery of singularities in general
relativity.

The consequences of our model in the ambit of cosmology are easily seen
by inspecting equations (\ref{93}) and (\ref{94}). The deformation of the
energy-momentum tensor in (\ref{94}) is written only in terms of the
noncommutative part $h^{\mu\nu}$. It would be interesting to study
whether or not $h^{\mu\nu}$ could account for a field of negative pressure.
If so, our model could describe the dynamics of dark energy.
Furthermore, the distortion of the gravitation expansion of the universe
due to the non-commutative degrees of freedom of the gravitational field
will certainly have some
effect on the anisotropy of the cosmic background
radiation, nucleosynthesis and structure formation. Of course, a detailed
analysis of these effects requires a careful study of the fluctuations
in the early universe.

Of course the validity
of this statements require further investigations.
The ultimate goal of this theory is to construct a consistent theory of
the gravitational field which is compatible with the Standard Model and able
to solve the current open issues which afflict General Relativity, i.e. the problems
of the origin of dark matter and dark energy, the recent anomalies found in
the solar system (Pioneer Anomaly, flyby anomaly, etc.) and last, but not least,
the problem of quantization of the gravitational field.

In summary, we would like to stress that our model makes it possible to make
a number of {\it very specific predictions}
that can serve as experimental tests of the theory.

%=============================
\section*{Acknowledgements}

I.G.A. would like to thank Frank Meyer for the discussion of the
paper \cite{aschieri05}.

%=================================================================
\appendix
\section*{Appendix A. Coincidence Limits}
\setcounter{equation}0
\renewcommand\theequation{A.\arabic{equation}}

In this appendix we will list the various coincidence limits used
during the calculations performed in this paper.

Through all this appendix the subscripts $0$,$1$ and $2$ will be
used to denote terms of different order in the deformation
parameter $\lambda$. More precisely for any quantity $\mathcal{X}$
which contains different orders of $\lambda$ we write
\begin{displaymath}
\mathcal{X}=\mathcal{X}_{0}+\lambda\mathcal{X}_{1}
+\lambda^{2}\mathcal{X}_{2}+O(\lambda^{3})\;,
\end{displaymath}
where $\mathcal{X}_{0}$, $\mathcal{X}_{1}$ and $\mathcal{X}_{2}$
denote, respectively, the zeroth, first and second order in
$\lambda$.

We start with the coincidence limit of the operator $\tilde{H}$ in
(\ref{30}) and its derivatives. More precisely we have
\begin{equation}\label{A1}
[\tilde{H}]=\lambda\xi_{\mu}\xi_{\nu}h^{\mu\nu}\;.
\end{equation}
For the first derivative we obtain
\begin{equation}\label{A2}
[\tilde{H}_{;\;\mu}]=\lambda\xi_{\alpha}\xi_{\beta}h^{\alpha\beta}{}_{;\;\mu}\;.
\end{equation}
For the second derivative we get the following formula
\begin{equation}\label{A3}
[\tilde{H}_{;\;\mu\nu}]=-\frac{2}{3}
\xi_{\alpha}\xi^{\rho}R^{\alpha}{}_{\mu\nu\rho}
-\frac{2}{3}\lambda\xi_{\alpha}\xi_{\sigma}h^{\rho\sigma}
(R^{\alpha}{}_{\mu\nu\rho}+R^{\alpha}{}_{\rho\nu\mu})
+\lambda\xi_{\alpha}\xi_{\beta}h^{\alpha\beta}{}_{;\;\mu\nu}\;.
\end{equation}

Recall, now, the definition (\ref{61a}) for the operator $K$. The
coincidence limits of the terms in $K$ are
\begin{eqnarray}\label{A4}
\left[B^{\rho^{\prime}\nu}_{0}\right]&=&-2g^{\rho\nu}\;,\\
\left[B^{\rho^{\prime}\nu}_{1}\right]&=&-2h^{\rho\nu}\;.
\end{eqnarray}
For the derivatives of these quantities we have
\begin{eqnarray}\label{A5}
\left[(\nabla_{\mu}B^{\rho^{\prime}\nu})_{0}\right]&=&0\;,\\
\left[(\nabla_{\mu}B^{\rho^{\prime}\nu})_{1}\right]&=&
-2h^{\rho\nu}{}_{;\;\mu}\;.
\end{eqnarray}
For the other terms we have
\begin{eqnarray}\label{A6}
\left[G^{\rho^{\prime}}_{0}\right]&=&0\;,\\
\left[G^{\rho^{\prime}}_{1}\right]&=&-h^{\rho\nu}{}_{;\;\nu}\;.
\end{eqnarray}
for the derivatives of (\ref{A6}) we get
\begin{eqnarray}\label{A7}
\left[(\nabla_{\nu}G^{\rho^{\prime}})_{0}\right]&=&
\frac{1}{3}R^{\rho}{}_{\nu}+\mathcal{R}^{\rho}{}_{\nu}\;,\\
\left[(\nabla_{\nu}G^{\rho^{\prime}})_{1}\right]&=&
-h^{\rho\alpha}{}_{;\;\alpha\nu}-\frac{1}{3}h^{\rho\alpha}
R_{\alpha\nu}+\frac{2}{3}h^{\mu\alpha}R^{\rho}{}_{\alpha\nu\mu}
+h^{\rho\alpha}\mathcal{R}_{\alpha\nu}\;,\\
\left[(\nabla_{\nu}G^{\rho^{\prime}})_{2}\right] &=&
-[\sigma_{;\;\alpha\nu},h^{\rho\alpha}]
-[\sigma_{;\;\alpha},h^{\rho\alpha}{}_{;\;\nu}]
-[\sigma^{;\;\rho}{}_{;\;\nu},\sigma]-[\sigma^{;\;\rho},\sigma_{;\;\nu}]\;.
\end{eqnarray}

For the operator $\mathcal{L}$ in (\ref{60}) we need the following
coincidence limits
\begin{eqnarray}\label{A8}
\left[(\mathcal{A}_{\nu})_{0}\right]&=&-[(\bar{\mathcal{A}}_{\nu})_{0}]
=-\phi_{;\;\nu}\;,\\
\left[(\mathcal{A}_{\nu})_{0}\right]&=&-[(\bar{\mathcal{A}}_{\nu})_{0}]
=-\sigma_{;\;\nu}\;,\\
\left[(\mathcal{A}_{\nu})_{0}\right]&=&[(\bar{\mathcal{A}}_{\nu})_{0}]
=\frac{1}{2}[\sigma_{;\;\nu},\sigma]\;.
\end{eqnarray}
We also used, during the calculation, the coincidence limits for
the derivatives of $\mathcal{A}_{\nu}$, namely
\begin{eqnarray}\label{A9}
\left[(\nabla_{\mu}\mathcal{A}_{\nu})_{0}\right]
&=&-\phi_{;\;\nu\mu}+\frac{1}{6}R_{\nu\mu}-\frac{1}{2}\mathcal{R}_{\nu\mu}\;,\\
\left[(\nabla_{\mu}\mathcal{A}_{\nu})_{1}\right]
&=&-\sigma_{\nu\mu}\;,\\
\left[(\nabla_{\mu}\mathcal{A}_{\nu})_{2}\right]
&=&\frac{1}{2}([\sigma_{;\;\nu\mu},\sigma]
+[\sigma_{;\;\nu},\sigma_{;\;\mu}])\;.
\end{eqnarray}

%=================================================================
\section*{Appendix B. Integrals}
\setcounter{equation}0
\renewcommand\theequation{B.\arabic{equation}}

In this appendix we will explicitly derive the solution to the
integrals in (\ref{70}) and (\ref{77}). The first integral that we
want to evaluate is the following
\begin{equation}\label{A10}
I_{1}(\alpha,\beta,\gamma,\delta) =\int\limits_{0}^{1}d\tau_{2}
\int\limits_{0}^{\tau_{2}}d\tau_{1}\tau_{1}^{\alpha}
(1-\tau_{2})^{\beta}(\tau_{2}-\tau_{1})^{\gamma}(1
-\tau_{2}+\tau_{1})^{\delta}\;.
\end{equation}
This integral is well defined for $\textrm{Re}(\alpha)>-1$,
$\textrm{Re}(\gamma)>-1$ and $0<\tau_{2}<1$. By using the integral
representation of the hypergeometric function we can evaluate the
integral in $\tau_{1}$, i.e.
\begin{equation}\label{A11}
I_{1}=\frac{\Gamma(1+\alpha)\Gamma(1+\gamma)}{\Gamma(2+\alpha
+\gamma)}\int\limits_{0}^{1}d\tau_{2}\tau_{2}^{1+\alpha
+\gamma}(1-\tau_{2})^{\beta+\delta}{}_{2}F_{1}
\left(1+\alpha,-\delta,2+\alpha+\gamma\;;\frac{\tau_{2}}{1-\tau_{2}}\right)\;.
\end{equation}
Now, by using the linear transformation formula for the
hypergeometric function we get \cite{abramowitz72}
\begin{eqnarray}\label{A12}
\lefteqn{{}_{2}F_{1}\left(1+\alpha,-\delta,2
+\alpha+\gamma\;;\frac{\tau_{2}}{1-\tau_{2}}\right)=}
\nonumber\\
&&=(1-\tau_{2})^{1+\alpha}{}_{2}F_{1}\left(1+\alpha,2
+\alpha+\gamma+\delta,2+\alpha+\gamma\;;\tau_{2}\right)\;.
\end{eqnarray}
By substituting the last expression in the integral (\ref{A11}) we
obtain
\begin{eqnarray}\label{A13}
\lefteqn{I_{1}=\frac{\Gamma(1+\alpha)\Gamma(1+\gamma)}{\Gamma(2
+\alpha+\gamma)}\times}\nonumber\\
&&\int\limits_{0}^{1}d\tau_{2}\tau_{2}^{1+\alpha
+\gamma}(1-\tau_{2})^{1+\alpha+\beta+\delta}{}_{2}F_{1}\left(1
+\alpha,2+\alpha+\gamma+\delta,2+\alpha+\gamma\;;\tau_{2}\right)\;.
\nonumber\\
&&\phantom{2}
\end{eqnarray}
This integral over $\tau_{2}$ is, now, of the following general
form
\begin{equation}\label{A14}
\mathcal{I}=\int\limits_{0}^{1}dx\;x^{c-1}(1-x)^{d-1}{}_{2}F_{1}
\left(a,b,c\;;x\right)\;,
\end{equation}
which is well define for $\textrm{Re}(c)>0$ and $\textrm{Re}(d)>0$
and has a solution in a closed form \cite{gradshtein07}, namely
\begin{equation}\label{A15}
\mathcal{I}=\frac{\Gamma(c)\Gamma(d)\Gamma(c+d-a-b)}{\Gamma(c+d-a)
\Gamma(c+d-b)}\;.
\end{equation}
By using the results (\ref{A15}) in the integral (\ref{A13}), we
get the solution (\ref{71}), i.e.
\begin{equation}\label{A16}
I_{1}=\frac{\Gamma(1+\alpha)\Gamma(1+\beta)\Gamma(1+\gamma)
\Gamma(2+\alpha+\beta+\delta)}{\Gamma(3+\alpha+\beta+\gamma+\delta)
\Gamma(2+\alpha+\beta)}
\end{equation}

The next integral that we used in our paper is the one in
(\ref{77}), namely
\begin{equation}\label{A17}
I_{2}(\alpha,\beta)=\int\limits_{0}^{1}d\tau_{1}\tau_{1}^{\alpha}(1-\tau_{1})^{\beta}\;.
\end{equation}
The solution to this integral is easily find by recalling the
integral representation of the hypergeometric function
\cite{abramowitz72}
\begin{equation}\label{A18}
{}_{2}F_{1}\left(a,b,c\;;z\right)=\frac{\Gamma(c)}{\Gamma(b)
\Gamma(c-b)}\int\limits_{0}^{1}dt\;t^{b-1}(1-t)^{c-b-1}(1-tz)^{-a}\;,
\end{equation}
where $\textrm{Re}(c)>0$ and $\textrm{Re}(b)>0$. From this last
general expression we obtain the integral (\ref{A17}) by setting
$z=0$, $\alpha=b-1$ and $\beta=c-b-1$. By recalling that
${}_{2}F_{1}\left(a,b,c\;;0\right)=1$, we finally get
\begin{equation}\label{A19}
I_{2}(\alpha,\beta)=\frac{\Gamma(1+\alpha)\Gamma(1+\beta)}{\Gamma(2
+\alpha+\beta)}=B(1+\alpha,1+\beta)\;,
\end{equation}
where $B(a,b)$ denotes the Euler beta function.

%============================================

\end{document}